\documentclass[12pt,a4paper]{article}               % I'm using a double-sided book style    

\usepackage{techrep_icg}
\usepackage{graphicx}
\usepackage{multirow}
\usepackage{xspace}
\usepackage{subfigure}
\usepackage{wrapfig}
\usepackage{xcolor}
\usepackage{morefloats}

\usepackage{multibib}
\newcites{links}{Weblinks}

\makeatletter
\DeclareRobustCommand\onedot{\futurelet\@let@token\@onedot}
\def\@onedot{\ifx\@let@token.\else.\null\fi\xspace}

\def\etal{\emph{et al}\onedot}

\makeatother

\begin{document}

%----------------------------------------------------------------------

\reportnr{2015-001}               % Number of the technical report

\title{The History of Mobile Augmented Reality} % Title of technical report

\subtitle{
Developments in Mobile AR over the last almost 50 years
} % Subtitle of technical report (use small letters only)

\repcity{Graz}            % City where the report was created

\repdate{\today}          % Date of creation

 % keywords that appear below the abstract

%----------------------------------------------------------------------

% List of authors

%

% List each author using a separate \author{} command

% If there is more than one author address, add a label to each author

% of the form \author[label]{name}.  This label should be identical to

% the corresponding label provided with the \address command

% N.B. It is not possible to link an author to more than one address.

%

\author[ICG]{Clemens Arth}
\author[ICG]{Lukas Gruber}
\author[ICG]{Raphael Grasset}
\author[ICG]{Tobias Langlotz}
\author[ICG]{Alessandro Mulloni}
\author[ICG]{Dieter Schmalstieg}
\author[ICG]{Daniel Wagner}

%----------------------------------------------------------------------

% List of addresses

%

% If there is more than one address, list each using a separate

% \address command using a label to link it to the respective author

% as described above

\newcommand{\TUGn}{Graz University of Technology}

\address[ICG]{Inst. for Computer Graphics and Vision \\ \TUGn, Austria}

%----------------------------------------------------------------------

% Information about the contact author

% if \contact is not defined (uncommented) or empty, the contact

%  information on the title page is suppressed.

% Name of contact

\contact{Clemens Arth}

% Email address of contact - do not use any LaTeX formatting here

\contactemail{arth@icg.tugraz.at}

%----------------------------------------------------------------------

% Do not alter the following line

\begin{abstract}
This document summarizes the major milestones in mobile Augmented Reality between 1968 and 2014. Mobile Augmented Reality has largely evolved over the last decade, as well as the interpretation itself of what is Mobile Augmented Reality. The first instance of Mobile AR can certainly be associated with the development of wearable AR, in a sense of experiencing AR during locomotion (mobile as a motion). With the transformation and miniaturization of physical devices and displays, the concept of mobile AR evolved towards the notion of "mobile device", aka AR on a mobile device. In this history of mobile AR we considered both definitions and the evolution of the term over time. 

Major parts of the list were initially compiled by the member of the Christian Doppler Laboratory for Handheld Augmented Reality in 2009 (author list in alphabetical order) for the ISMAR society. More recent work was added in 2013 and during preparation of this report.  

Permission is granted to copy and modify. Please email the first author if you find any errors.
\end{abstract}

\section*{Introduction}  % Article style

This document summarizes the major milestones in mobile Augmented Reality between 1968 and 2014. Mobile Augmented Reality has largely evolved over the last decade, as well as the interpretation itself of what is Mobile Augmented Reality. The first instance of Mobile AR can certainly be associated with the development of wearable AR, in a sense of experiencing AR during locomotion (mobile as a motion). With the transformation and miniaturization of physical devices and displays, the concept of mobile AR evolved towards the notion of "mobile device", aka AR on a mobile device. In this history of mobile AR we considered both definitions and the evolution of the term over time. 

Major parts of the list were initially compiled by the member of the The list was compiled by the member of the Christian Doppler Laboratory for Handheld Augmented Reality\footnote{CDL on Handheld AR: \url{http://studierstube.org/handheld_ar/}} in 2009 (author list in alphabetical order) for the ISMAR society. More recent work was added in 2013 and during preparation of this report.  

Permission is granted to copy and modify. Please email the first author if you find any errors.

\vspace{20pt}

\begin{figure}[htbp]
\centering
\begin{tabular}{p{90pt}p{90pt}p{90pt}p{90pt}}
\includegraphics[height=2.0cm]{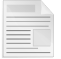} &
\includegraphics[height=2.0cm]{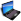} &
\includegraphics[height=2.0cm]{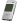} &
\includegraphics[height=2.0cm]{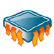} \\
\small (a) Research & \small (b) Mobile PC & \small (c) Mobile Phone & \small (d) Hardware \\
\vspace{0.1in} \\
\includegraphics[height=2.0cm]{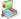} &
\includegraphics[height=2.0cm]{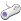} &
\includegraphics[height=2.0cm]{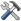} &
\includegraphics[height=2.0cm]{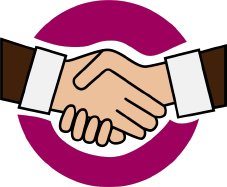} \\
\small (e) Standard & \small (f) Game & \small (g) Tool & (h) Deal
\end{tabular}
\caption{Icons used throughout this report for a rough categorization of related research, development and events.} \label{fig:zero}
\vspace{-20pt}
\end{figure}

\newpage

\section*{1968}
\begin{wrapfigure}{l}{2.1cm}
	\vspace{-10pt}	
	\includegraphics[width=1.0cm]{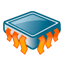}
	\includegraphics[width=1.0cm]{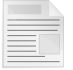}
	\vspace{-20pt}		
\end{wrapfigure}
\begin{figure}[tbp]
\vspace{-10pt}
\centering
\subfigure[]{\label{fig:firsta}\includegraphics[height=4.5cm]{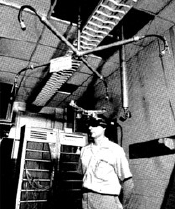}} \hfill
\subfigure[]{\label{fig:firstb}\includegraphics[height=4.5cm]{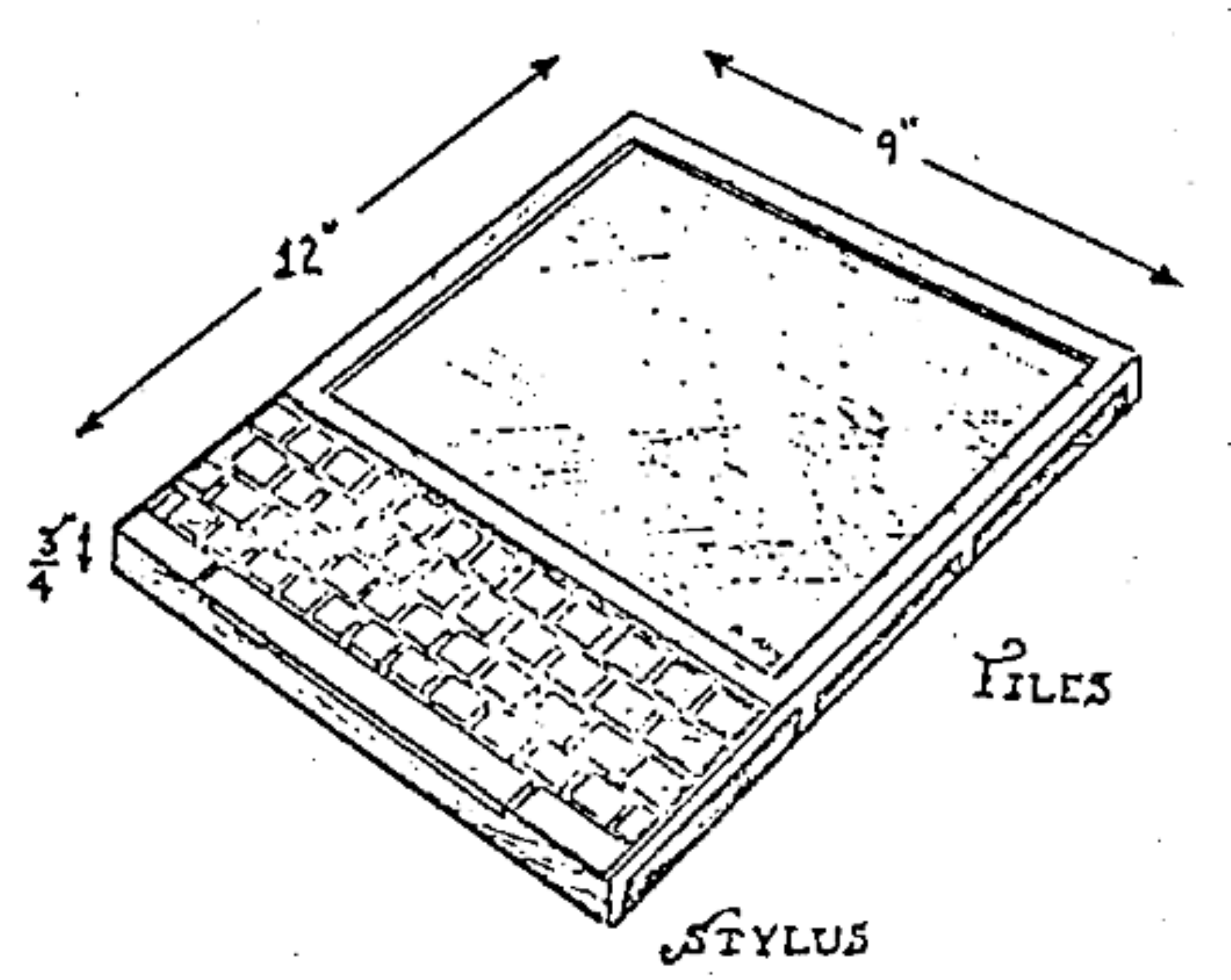}} \hfill
\subfigure[]{\label{fig:firstc}\includegraphics[height=4.5cm]{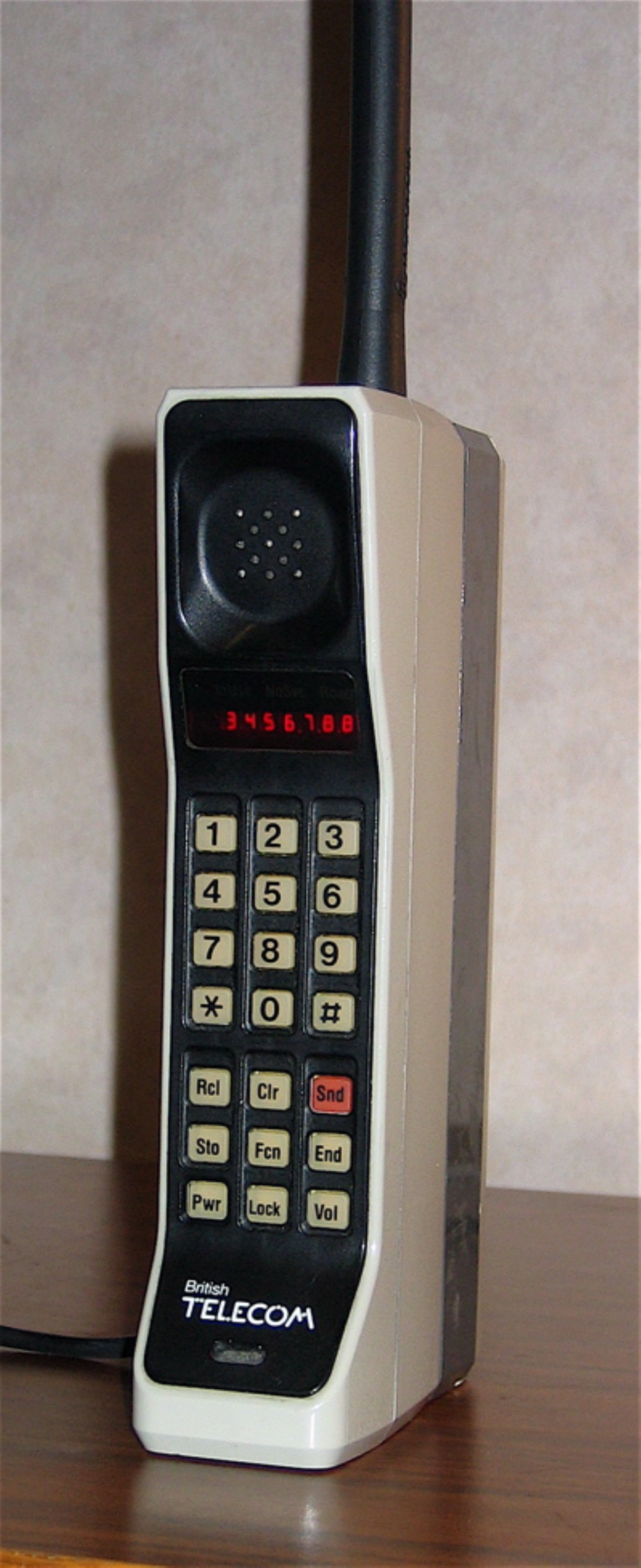}}
\subfigure[]{\label{fig:firstd}\includegraphics[height=4.5cm]{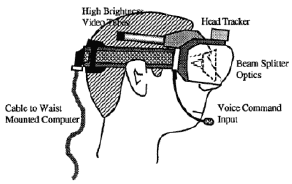}} \hfill
\subfigure[]{\label{fig:firste}\includegraphics[height=4.5cm]{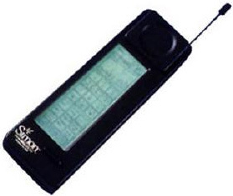}}
\vspace{-10pt}
\caption{(a): Sutherland's system in \cite{Sutherland1968}. (b): Conceptual Tablet Computer by Kay in 1972 \cite{Kay72}. (c): First handheld mobile phone by Motorola in 1973. (d): Caudell and Mizell coining AR in 1992 \cite{Caudell92}. (e): IBM smartphone presented in 1992.} \label{fig:first}
\end{figure}

Ivan Sutherland \cite{Sutherland1968} creates the \textbf{first augmented reality system}, which is also the first virtual reality system (see Fig.\ref{fig:firsta} left). It uses an optical see-through head-mounted display that is tracked by one of two different 6DOF trackers: a mechanical tracker and an ultrasonic tracker. Due to the limited processing power of computers at that time, only very simple wireframe drawings could be displayed in real time.

\vspace{-5pt}
\section*{1972}

\begin{wrapfigure}{L}{1.1cm}
	\vspace{-17pt}	
	\includegraphics[width=1.0cm]{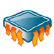}
	\vspace{-15pt}		
\end{wrapfigure}
The \textbf{first conceptual tablet computer} was proposed in 1972 by Alan Kay, named the Dynabook \cite{Kay72}. The Dynabook was proposed as personal computer for children, having the format factor of a tablet with a mechanical keyboard (really similar design from the One Laptop per Child project started in 2005). The Dynabook is probably recognized as being the precursor of the tablet computers decades before the iPad (see Fig. \ref{fig:firstb}).

\vspace{-5pt}
\section*{1973}

\begin{wrapfigure}{L}{1.1cm}
	\vspace{-17pt}	
	\includegraphics[width=1.0cm]{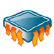}
	\vspace{-15pt}		
\end{wrapfigure}
The \textbf{first handheld mobile phone} was presented by Motorola and demonstrated in April 1973 by Dr Martin Cooper \cite{Cooper73}. The mobile named DynaTAC for Dynamic Adaptive Total Area Coverage was supporting only 35 minutes of call (see Fig. \ref{fig:firstc}).

\section*{1982}
\begin{wrapfigure}{L}{1.1cm}
	\vspace{-10pt}	
	\includegraphics[width=1.0cm]{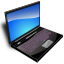}
	\vspace{-10pt}		
\end{wrapfigure}
The \textbf{first laptop}, the Grid Compass\footnote{\url{http://home.total.net/~hrothgar/museum/Compass/}} 
1100 is released, which was also the first computer to use a clamshell design. The Grid Compass 1100 had an Intel 8086 CPU, 350 Kbytes of memory and a display with a resolution of 320x240 pixels, which was extremely powerful for that time and justified the enormous costs of 10.000 USD. However, its weight of 5kg made it hardly portable.

\vspace{-5pt}
\section*{1992}
\begin{wrapfigure}{L}{1.1cm}
	\vspace{-10pt}	
	\includegraphics[width=1.0cm]{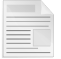}
	\vspace{-10pt}		
\end{wrapfigure}
Tom Caudell and David Mizell coin the term \textbf{"augmented reality"} to refer to overlaying computer-presented material on top of the real world \cite{Caudell92} (see Fig.\ref{fig:firstd}). Caudell and Mizell discuss the advantages of AR versus VR such as requiring less processing power since less pixels have to be rendered. They also acknowledge the increased registration requirements in order to align real and virtual.

\vspace{0.1in}

\begin{wrapfigure}{L}{1.1cm}
	\vspace{-10pt}	
	\includegraphics[width=1.0cm]{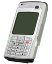}
	\vspace{-10pt}		
\end{wrapfigure}
\noindent At COMDEX 1992, IBM and Bellsouth introduce the \textbf{first smartphone}, the IBM Simon Personal Communicator\footnote{Wikipedia: \url{http://en.wikipedia.org/wiki/Simon_(phone)}}, which was released in 1993 (see Fig.\ref{fig:firste}). The phone has 1 Megabyte of memory and a B/W touch screen with a resolution of 160 x 293 pixels. The IBM Simon works as phone, pager, calculator, address book, fax machine, and e-mail device. It weights 500 grams and cost 900 USD.

\newpage

\vspace{-5pt}
\section*{1993}
\begin{wrapfigure}{L}{2.1cm}
	\vspace{-10pt}	
	\includegraphics[width=1.0cm]{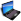}
	\includegraphics[width=1.0cm]{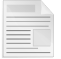}
	\vspace{-20pt}		
\end{wrapfigure}
Loomis \etal develop a prototype of an \textbf{outdoor navigation system for visually impaired} \cite{Loomis93}. They combine a notebook with a differential GPS receiver and a head-worn electronic compass. The application uses data from a GIS (Geographic Information System) database and provides navigational assistance using an "acoustic virtual display": labels are spoken using a speech synthesizer and played back at correct locations within the auditory space of the user.

\begin{figure}
\vspace{-10pt}
\centering
\subfigure[]{\label{fig:seconda}\includegraphics[height=3.3cm]{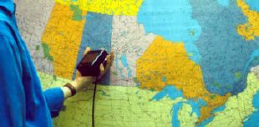}} \hfill
\subfigure[]{\label{fig:secondb}\includegraphics[height=3.3cm]{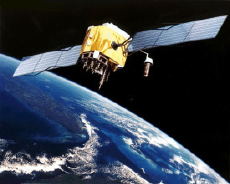}}
\subfigure[]{\label{fig:secondc}\includegraphics[height=3.3cm]{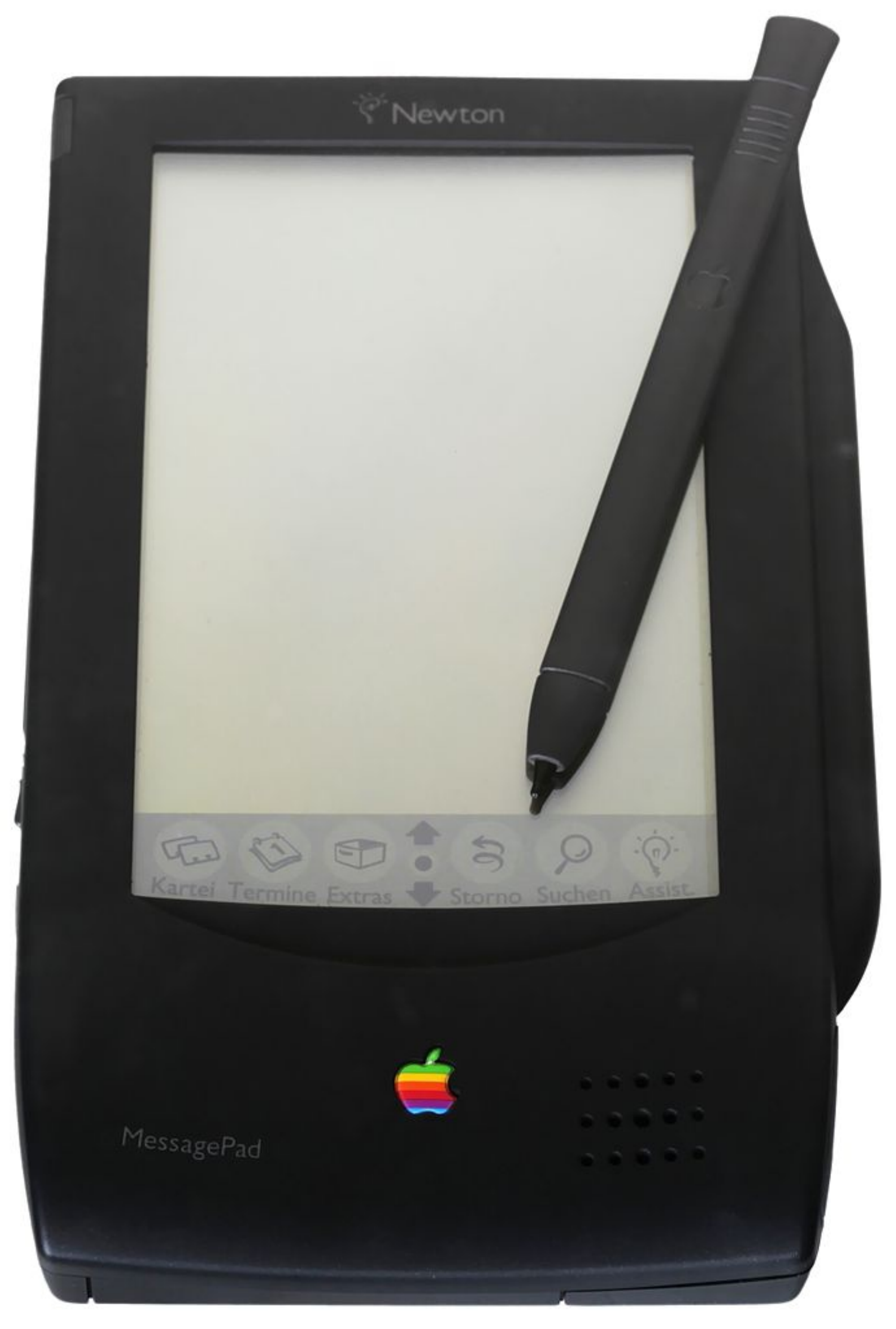}}
\vspace{-10pt}
\caption{(a): Chameleon system proposed by Fitzmaurice \cite{Fitzmaurice93}. (b): NAVSTAR-GPS goes live in 1993. (c): Apple Newton Message Pad 100.} \label{fig:second}
\end{figure}

\vspace{0.1in}

\begin{wrapfigure}{L}{2.1cm}
	\vspace{-10pt}	
	\includegraphics[width=1.0cm]{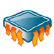}
	\includegraphics[width=1.0cm]{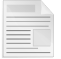}
	\vspace{-20pt}		
\end{wrapfigure}
\noindent Fitzmaurice creates \textbf{Chameleon} (see Fig.\ref{fig:seconda}), a key example of displaying spatially situated information with a tracked hand-held device. In his setup the output device consists of a 4" screen connected to a video camera via a cable \cite{Fitzmaurice93}. The video camera records the content of a Silicon Graphics workstation's large display in order to display it on the small screen. Fitzmaurice uses a tethered magnetic tracker (Ascension bird) for registration in a small working environment. Several gestures plus a single button allow the user to interact with the mobile device. Chameleon's mobility was strongly limited due to the cabling. It did also not augment in terms of overlaying objects on a video feed of the real world. 

\vspace{0.1in}

\begin{wrapfigure}{L}{1.1cm}
	\vspace{-10pt}	
	\includegraphics[width=1.0cm]{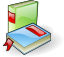}
	\vspace{-10pt}		
\end{wrapfigure}
\noindent In December 1993 the \textbf{Global Positioning System} (GPS, official name "NAVSTAR-GPS") achieves initial operational capability (see Fig.\ref{fig:secondb}). Although GPS\footnote{Wikipedia: \url{http://en.wikipedia.org/wiki/Global_Positioning_System}} was originally launched as a military service, nowadays millions of people use it for navigation and other tasks such as geo-caching or Augmented Reality. A GPS receiver calculates its position by carefully timing the signals sent by the constellation of GPS satellites. The accuracy of civilian GPS receivers is typically in the range of 15 meter. More accuracy can be gained by using Differential GPS (DGPS) that uses correction signals from fixed, ground-based reference stations.

\vspace{0.1in}

\begin{wrapfigure}{L}{1.1cm}
	\vspace{-17pt}	
	\includegraphics[width=1.0cm]{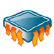}
	\vspace{-15pt}		
\end{wrapfigure}
\noindent The Apple Newton Message Pad 100 was one of the earliest \textbf{commercial personal digital assistant (PDA)}\footnote{Wikipedia: \url{http://en.wikipedia.org/wiki/MessagePad}}. Equipped with a stylus and handwritten recognition, and feature a screen in black and white of 336x240 pixels (see Fig. \ref{fig:secondc}).

\vspace{-5pt}
\section*{1994}
\begin{wrapfigure}{L}{1.1cm}
	\vspace{-17pt}	
	\includegraphics[width=1.0cm]{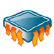}
	\vspace{-15pt}		
\end{wrapfigure}
Steve Mann starts \textbf{wearing a webcam for almost 2 years}. From 1994-1996 Mann wore a mobile camera plus display for almost every waking minute. Both devices were connected to his website allowing online visitors to see what Steve was seeing and to send him messages that would show up on his mobile display\footnote{S. Mann, “Wearable Wireless Webcam,” personal WWW page. \url{wearcam.org}}

\begin{figure}[tbp]
\centering
\vspace{-50pt}
\subfigure[]{\label{fig:thirda}\includegraphics[height=1.8cm]{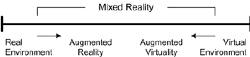}} \hfill
\subfigure[]{\label{fig:thirdb}\includegraphics[height=3.5cm]{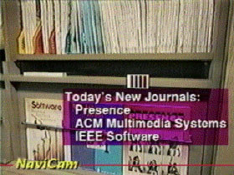}} \\
\subfigure[]{\label{fig:thirdc}\includegraphics[height=4.4cm]{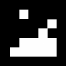}} \hfill
\subfigure[]{\label{fig:thirdd}\includegraphics[height=4.4cm]{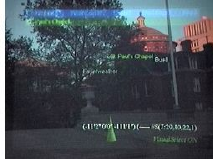}} \hfill
\subfigure[]{\label{fig:thirde}\includegraphics[height=4.4cm]{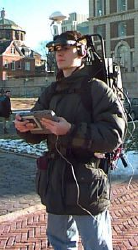}}
\vspace{-10pt}
\caption{(a): Milgram Continuum \cite{Milgram94}. (b): Rekimotos NaviCam system \cite{Rekimoto95}. (c): Rekimoto's matrix marker \cite{Rekimoto96}. (d) and (e): Touring Machine by Feiner \etal \cite{Feiner97}.} \label{fig:third}
\end{figure}

\vspace{0.1in}

\begin{wrapfigure}{L}{1.1cm}
	\vspace{-10pt}	
	\includegraphics[width=1.0cm]{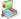}
	\vspace{-15pt}		
\end{wrapfigure}
\noindent Paul Milgram and Fumio Kishino write their seminal paper "Taxonomy of Mixed Reality Visual Displays" in which they define the \textbf{Reality-Virtuality Continuum} \cite{Milgram94} (see Fig.\ref{fig:thirda}). Milgram and Kishino describe a continuum that spans from the real environment to the virtual environment. In between there are Augmented Reality, closer to the real environment and Augmented Virtuality, which is closer to the virtual environment. Today Milgram's Continuum and Azuma's definition (1997) are commonly accepted as defining Augmented Reality.

\vspace{-5pt}
\section*{1995}
\begin{wrapfigure}{L}{2.1cm}
	\vspace{-10pt}	
	\includegraphics[width=1.0cm]{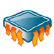}
	\includegraphics[width=1.0cm]{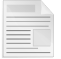}	
	\vspace{-20pt}		
\end{wrapfigure}
Jun Rekimoto and Katashi Nagao create the NaviCam, a tethered setup, similar to Fitzmaurice's Chameleon \cite{Rekimoto95} (see Fig.\ref{fig:thirdb}). The NaviCam also uses a nearby powerful workstation, but has a camera mounted on the mobile screen that is used for optical tracking. The computer detects color-coded markers in the live camera image and displays context sensitive information directly on top of the video feed in a see-through manner.

\vspace{0.1in}

\begin{wrapfigure}{L}{1.1cm}
	\vspace{-10pt}
	\includegraphics[width=1.0cm]{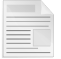}
	\vspace{-10pt}		
\end{wrapfigure}
\noindent Benjamin Bederson introduced the term \textbf{Audio Augmented Reality} by presenting a system that demonstrated an augmentation of the audition modality \cite{Bederson95}. The developed prototype uses a MD-player which plays audio information based on the tracked position of the user as part of a museum guide.

 \vspace{-5pt}
\section*{1996}
\begin{wrapfigure}{L}{1.1cm}
	\vspace{-10pt}
	\includegraphics[width=1.0cm]{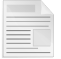}
	\vspace{-10pt}		
\end{wrapfigure}
Jun Rekimoto presents 2D matrix markers\footnote{\url{http://www.sonycsl.co.jp/person/rekimoto/matrix/Matrix.html}} (square-shaped barcodes), one of the first marker systems to allow camera tracking with six degrees of freedom \cite{Rekimoto96} (see Fig.\ref{fig:thirdc}).

\section*{1997}
\begin{wrapfigure}{L}{1.1cm}
	\vspace{-10pt}	
	\includegraphics[width=1.0cm]{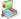}
	\vspace{-10pt}		
\end{wrapfigure}
Ronald Azuma presents the \textbf{first survey on Augmented Reality}. In his publication, Azuma provides a widely acknowledged definition for AR \cite{Azuma97}, as identified by three characteristics:

\begin{itemize}
\item it combines real and virtual
\item it is interactive in real time
\item it is registered in 3D.
\end{itemize}
\vspace{0.1in} 

\begin{wrapfigure}{L}{2.1cm}
	\vspace{-10pt}	
	\includegraphics[width=1.0cm]{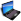}
	\includegraphics[width=1.0cm]{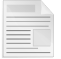}	
	\vspace{-20pt}		
\end{wrapfigure}
\noindent Steve Feiner \etal present the \textbf{Touring Machine}, the first mobile augmented reality system (\textbf{MARS}) \cite{Feiner97} (see Fig.\ref{fig:thirdd} and Fig. \ref{fig:thirde}). It uses a see-through head-worn display with integral orientation tracker; a backpack holding a computer, differential GPS, and digital radio for wireless web access; and a hand-held computer with stylus and touchpad interface\footnote{MARS: \url{http://graphics.cs.columbia.edu/projects/mars/mars.html}}.

\vspace{0.1in} 

\begin{wrapfigure}{L}{2.1cm}
	\vspace{0pt}	
	\includegraphics[width=1.0cm]{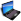}
	\includegraphics[width=1.0cm]{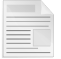}	
	\vspace{-20pt}		
\end{wrapfigure}
\noindent Thad Starner \etal explore possible applications of mobile augmented reality, creating a small community of users equipped with wearable computers interconnected over a network \cite{Starner97}. The explored applications include an information system for offices, people recognition and coarse localization with infrared beacons.

\vspace{0.1in}

\begin{wrapfigure}{L}{1.1cm}
	\vspace{-10pt}	
	\includegraphics[width=1.0cm]{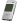}
	\vspace{-20pt}		
\end{wrapfigure}
\noindent Philippe Kahn invents the \textbf{camera phone}\footnote{Wikipedia Camera Phone: \url{http://en.wikipedia.org/wiki/Camera_phone}}, a mobile phone which is able to capture still photographs (see Fig.\ref{fig:fortha}). Back in 1997, Kahn used his invention to share a picture of his newborn daughter with more than 2000 relatives and friends, spread around the world. Today more than half of all mobile phones in use are camera phones.

\vspace{0.1in}

\begin{wrapfigure}{L}{1.1cm}
	\vspace{-17pt}	
	\includegraphics[width=1.0cm]{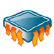}
	\vspace{-15pt}	
\end{wrapfigure}
\noindent Sony releases the Glasstron, a series of \textbf{optical HMD (optionally see-through) for the general public}. Adoption was rather small, but the affordable price of the HMD made it really popular in AR research labs and for the development of wearable AR prototype (see Fig. \ref{fig:forthb}).

\vspace{-5pt}
\section*{1998}

\begin{figure}[tbp]
\centering
\vspace{-50pt}
\subfigure[]{\label{fig:fortha}\includegraphics[height=3.9cm]{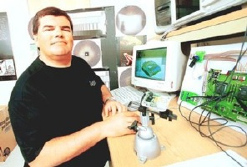}} \hfill
\subfigure[]{\label{fig:forthb}\includegraphics[height=3.9cm]{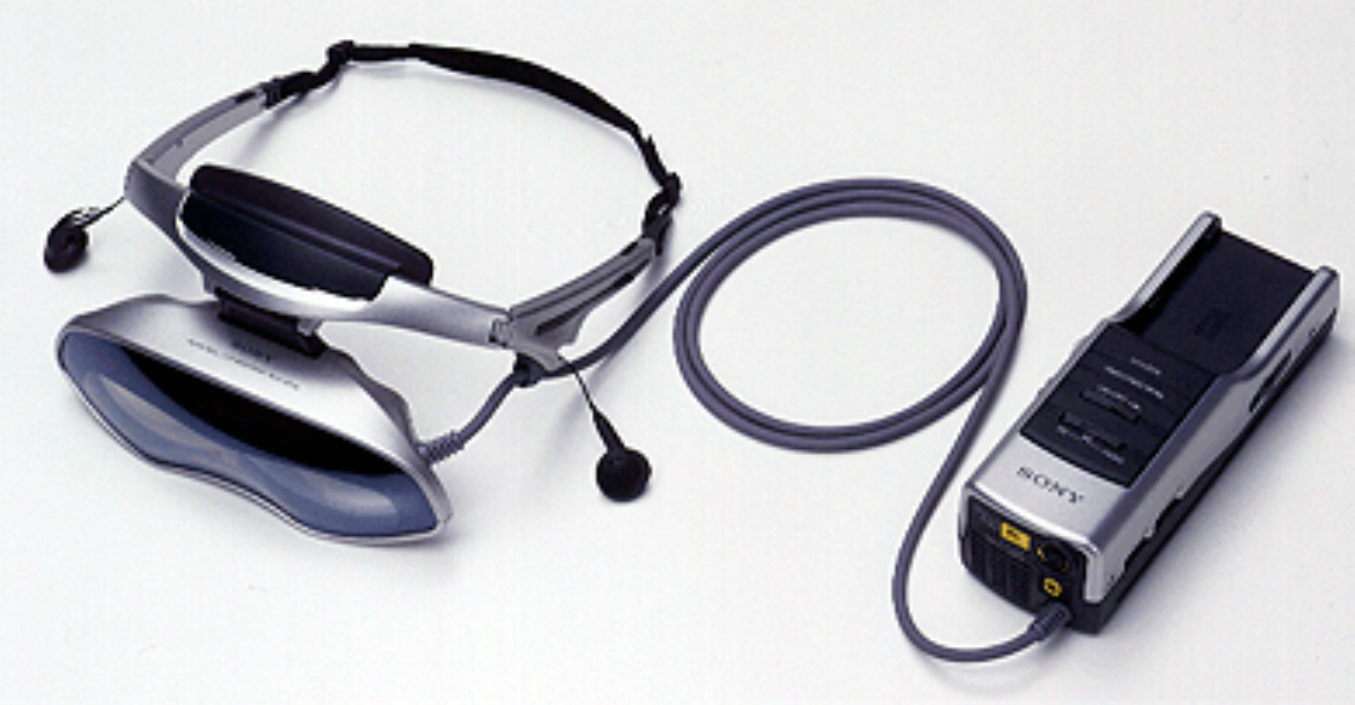}} \\
\subfigure[]{\label{fig:forthc}\includegraphics[height=4.5cm]{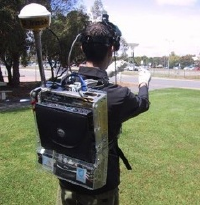}} \hfill
\subfigure[]{\label{fig:forthd}\includegraphics[height=4.5cm]{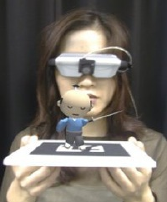}} \hfill
\subfigure[]{\label{fig:forthe}\includegraphics[height=4.5cm]{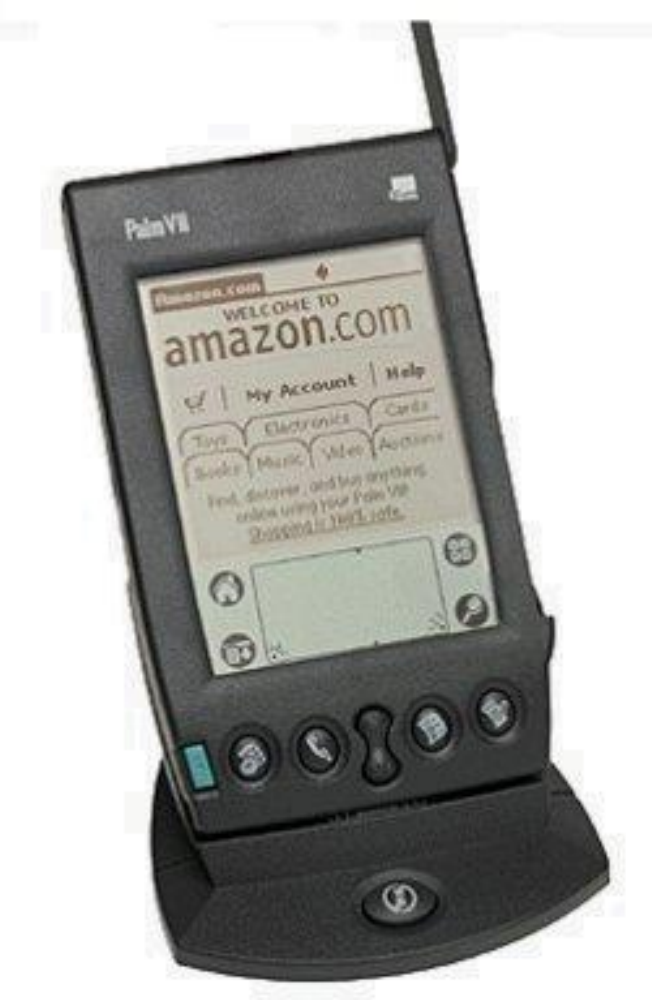}} 
\vspace{-10pt}
\caption{(a): Camera Phone Development by Kahn. (b): Sony Glasstron optical HMD in 1997. (c): Thomas \etal's Tinmith system \cite{Thomas98}. (d): ARToolKit for pose tracking in 6DOF \cite{Kato99}. (e): Palm VII, the first consumer LBS device.} \label{fig:forth}
\vspace{-10pt}
\end{figure}

\begin{wrapfigure}{L}{2.1cm}
	\vspace{-10pt}	
	\includegraphics[width=1.0cm]{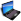}
	\includegraphics[width=1.0cm]{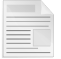}	
	\vspace{-20pt}		
\end{wrapfigure}
Bruce Thomas et al. present "Map-in-the-hat", a backpack-based wearable computer that includes GPS, electronic compass and a head-mounted display \cite{Thomas98} (see Fig.\ref{fig:forthc}). At this stage the system was utilized for navigation guidance, but it later evolved into \textbf{Tinmith}, an AR platform used for several other AR projects\footnote{Tinmith webpage: \url{http://www.tinmith.net/}}.

\vspace{-5pt}
\section*{1999}

\begin{wrapfigure}{L}{1.1cm}
	\vspace{-10pt}	
	\includegraphics[width=1.0cm]{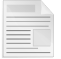}
	\vspace{-15pt}		
\end{wrapfigure}
Hirokazu Kato and Mark Billinghurst present \textbf{ARToolKit}, a pose tracking library with six degrees of freedom, using square fiducials and a template-based approach for recognition \cite{Kato99}. ARToolKit is available as open source under the GPL license and is still very popular in the AR community (see Fig. \ref{fig:forthd}).

\vspace{0.1in}

\begin{wrapfigure}{L}{2.1cm}
	\vspace{-0pt}	
	\includegraphics[width=1.0cm]{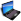}
	\includegraphics[width=1.0cm]{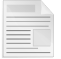}	
	\vspace{-20pt}		
\end{wrapfigure}
\begin{figure}[tbp]
\centering
\vspace{-50pt}
\subfigure[]{\label{fig:fiftha}\includegraphics[height=3.9cm]{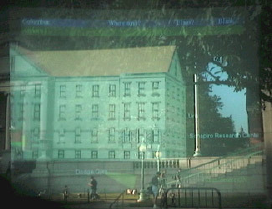}} \hfill
\subfigure[]{\label{fig:fifthb}\includegraphics[height=3.9cm]{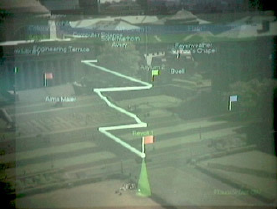}} \hfill
\subfigure[]{\label{fig:fifthc}\includegraphics[height=3.9cm]{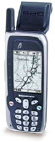}} 
\vspace{-10pt}
\caption{(a): H\"ollerer \etal's MARS system \cite{Hollerer99}. (b): H\"ollerer \etal's user interface \cite{Hollerer99b}. (c) Benfon Esc! NT2002, the first GSM phone with a built-in GPS sensor.} \label{fig:fifth}
\end{figure}

\noindent Tobias H\"ollerer \etal develop a \textbf{mobile AR} system that allows the user to explore hypermedia news stories that are located at the places to which they refer and to receive a guided campus tour that overlays models of earlier buildings \cite{Hollerer99} (see Fig. \ref{fig:fiftha}). This was the first mobile AR system to use RTK GPS and an inertial-magnetic orientation tracker.

\vspace{0.1in}

\begin{wrapfigure}{L}{2.1cm}
	\vspace{-15pt}	
	\includegraphics[width=1.0cm]{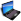}
	\includegraphics[width=1.0cm]{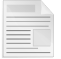}	
	\vspace{-20pt}		
\end{wrapfigure}
\noindent Tobias H\"ollerer \etal present a mobile augmented reality system that includes indoor user interfaces (desktop, AR tabletop, and head-worn VR) to interact with the outdoor user \cite{Hollerer99b} (see Fig. \ref{fig:fifthb}). While outdoor users experience a first-person spatialized multimedia presentation via a head-mounted display, indoor users can get an overview of the outdoor scene.

\vspace{0.1in}
\begin{wrapfigure}{L}{1.1cm}
	\vspace{-10pt}	
	\includegraphics[width=1.0cm]{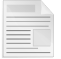}
	\vspace{-10pt}		
\end{wrapfigure}
\noindent Jim Spohrer publishes the \textbf{Worldboard} concept, a scalable infrastructure to support mobile applications that span from low-end location-based services, up to high-end mobile AR \cite{Spohrer99}. In his paper, Spohrer also envisions possible application cases for mobile AR, and social implications.

\newpage

\begin{wrapfigure}{L}{1.1cm}
	\vspace{-5pt}	
	\includegraphics[width=1.0cm]{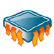}
	\vspace{-15pt}	
\end{wrapfigure}
\noindent The \textbf{first consumer LBS device} was the Palm VII, only supporting zip code based location services (see Fig.\ref{fig:forthd}). 2 years later, different mobile operators provided different location based services using private network technology\footnote{Wikipedia: \url{http://en.wikipedia.org/wiki/Palm_VII}}. 

\vspace{0.1in}

\begin{wrapfigure}{L}{1.1cm}
	\vspace{-10pt}	
	\includegraphics[width=1.0cm]{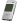}
	\vspace{-20pt}		
\end{wrapfigure}
\noindent Benefon Esc! NT2002\footnote{\url{http://www.benefon.de/products/esc/}}, the \textbf{first GSM phone with a built-in GPS receiver} is released in late 1999 (see Fig. \ref{fig:fifthc}). It had a black and white screen with a resolution of 100x160 pixels. Due to limited storage, the phone downloaded maps on demand. The phone also included a friend finder that exchanged GPS positions with other Esc! devices via SMS.

\vspace{0.1in}

\begin{wrapfigure}{L}{1.1cm}
	\vspace{-10pt}	
	\includegraphics[width=1.0cm]{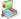}
	\vspace{-0pt}		
\end{wrapfigure}
\noindent The wireless network protocols 802.11a/802.11b\footnote{Wikipedia: \url{http://en.wikipedia.org/wiki/802.11}} - commonly known as \textbf{WiFi} - are defined. The original version - obsolete - specifies bitrates of 1 or 2 megabits per second (Mbit/s), plus forward error correction code.
 
\vspace{-5pt}
\section*{2000}

\begin{wrapfigure}{L}{3.2cm}
	\vspace{-10pt}	
	\includegraphics[width=1.0cm]{figs/icons/notebook}
	\includegraphics[width=1.0cm]{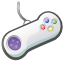}
	\includegraphics[width=1.0cm]{figs/icons/paper}		
	\vspace{-20pt}		
\end{wrapfigure}
\begin{figure}[tbp]
\centering
\vspace{-50pt}
\subfigure[]{\label{fig:sixa}\includegraphics[height=5.0cm]{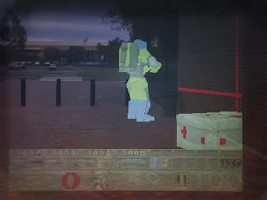}} \hfill
\subfigure[]{\label{fig:sixb}\includegraphics[height=5.0cm]{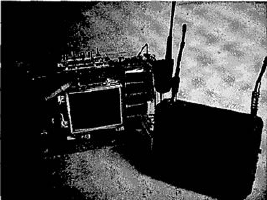}} \hfill
\subfigure[]{\label{fig:sixc}\includegraphics[height=3.5cm]{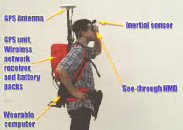}} \hfill
\subfigure[]{\label{fig:sixd}\includegraphics[height=3.5cm]{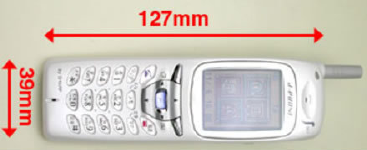}} 
\vspace{-10pt}
\caption{(a): ARQuake by Thomas \etal \cite{Thomas00}. (b): mPARD system by Regenbrecht and Specht \cite{Regenbrecht00}. (c): BARS system by Julier \etal \cite{Julier00}. (d): First commercial camera phone in 2000.} \label{fig:six}
\end{figure}

Bruce Thomas \etal present \textbf{AR-Quake}, an extension to the popular desktop game Quake \cite{Thomas00} (see Fig. \ref{fig:sixa}). ARQuake is a first-person perspective application which is based on a 6DOF tracking system using GPS, a digital compass and vision-based tracking of fiducial markers. Users are equipped with a wearable computer system in a backpack, an HMD and a simple two-button input device. The game can be played in- or outdoors where the usual keyboard and mouse commands for movement and actions are performed by movements of the user in the real environment and using the simple input interface.

\vspace{0.1in}

\begin{wrapfigure}{L}{2.1cm}
	\vspace{0pt}	
	\includegraphics[width=1.0cm]{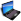}
	\includegraphics[width=1.0cm]{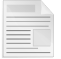}	
	\vspace{-20pt}		
\end{wrapfigure}
\noindent
Regenbrecht and Specht present \textbf{mPARD}, using analogue wireless video transmission to a host computer which is taking the burden of computation off the mobile hardware platform \cite{Regenbrecht00} (see Fig. \ref{fig:sixb}). The rendered and augmented images are sent back to the visualization device over a separate analog channel. The system can operate within 300m outdoors and 30m indoors, and the batteries allow for an uninterrupted operation of 5 hours at max.

\vspace{0.1in}

\begin{wrapfigure}{L}{1.1cm}
	\vspace{-10pt}	
	\includegraphics[width=1.0cm]{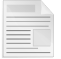}
	\vspace{-10pt}		
\end{wrapfigure}
\noindent Fritsch \etal introduce a general \textbf{architecture for large scale AR system} as part of the \textbf{NEXUS project}. The NEXUS model introduces the notion of augmented world using distributed data management and a variety of sensor system \cite{Fritsch01}.

\vspace{0.1in}

\begin{wrapfigure}{L}{2.1cm}
	\vspace{-10pt}	
	\includegraphics[width=1.0cm]{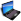}
	\includegraphics[width=1.0cm]{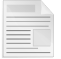}	
	\vspace{-20pt}		
\end{wrapfigure}
\noindent Simon Julier \etal present \textbf{BARS}, the Battlefield Augmented Reality System \cite{Julier00} (see Fig. \ref{fig:sixc}). The system consists of a wearable computer, a wireless network system and a see-through HMD. The system targets the augmentation of a battlefield scene with additional information about environmental infrastructure, but also about possible enemy ambushes.

\vspace{0.1in}

\begin{wrapfigure}{L}{1.1cm}
	\vspace{-0pt}	
	\includegraphics[width=1.0cm]{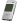}
	\vspace{-20pt}		
\end{wrapfigure}
\noindent Sharp corporation releases the \textbf{first commercial camera phone} to public (see Fig. \ref{fig:sixd}). The official name of the phone is J-SH04\footnote{\url{http://k-tai.impress.co.jp/cda/article/showcase_top/3913.html}}. The phones' camera has a resolution of 0.1 megapixels.

\vspace{0.1in}

\begin{wrapfigure}{L}{1.1cm}
	\vspace{-10pt}	
	\includegraphics[width=1.0cm]{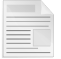}
	\vspace{-10pt}		
\end{wrapfigure}
\noindent At ISAR, Julier \etal described the problem of information overload and visual clutter within mobile Augmented Reality \cite{Julier02}. They proposed information filtering for mobile AR based on techniques such as physically-based methods, methods using the spatial model of interaction, rule-based filtering, and a combination of these methods to reduce the information overload in mobile AR scenarios.

\begin{figure}[tbp]
\centering
\vspace{-10pt}
\subfigure[]{\label{fig:sevena}\includegraphics[height=4.8cm]{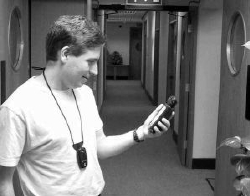}} \hfill
\subfigure[]{\label{fig:sevenb}\includegraphics[height=4.8cm]{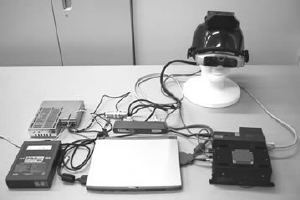}} \\
\subfigure[]{\label{fig:sevenc}\includegraphics[height=3.5cm]{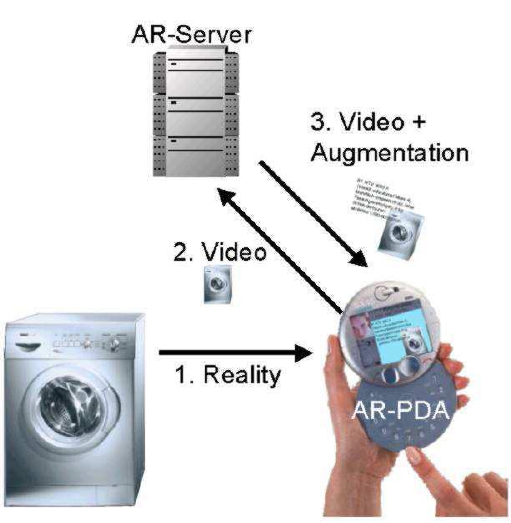}} \hfill
\subfigure[]{\label{fig:sevend}\includegraphics[height=3.5cm]{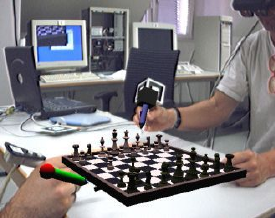}} \hfill
\subfigure[]{\label{fig:sevene}\includegraphics[height=3.5cm]{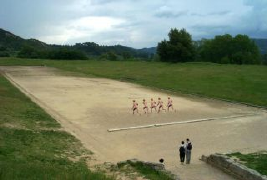}} \\
\subfigure[]{\label{fig:sevenf}\includegraphics[height=4.8cm]{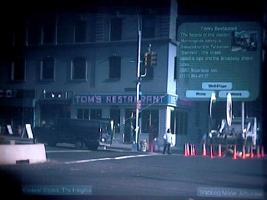}} \hfill
\subfigure[]{\label{fig:seveng}\includegraphics[height=4.8cm]{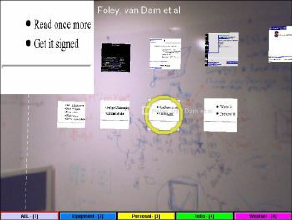}} 
\vspace{-10pt}
\caption{(a): BatPortal by Newman \etal \cite{Newman01}. (b): TOWNWEAR system by Hara \etal \cite{Satoh01}. (c): Wireless AR setup concept by Fruend \etal \cite{Fruend01}. (d): Multi-user AR system by Reitmayr and Schmalstieg \cite{Reitmayr01}. (e): ARCHEOGUIDE by Flahakis \etal \cite{Vlahakis2001}. (f): Mobile AR restaurant guide by Bell \etal \cite{Bell01}. (g): First AR browser by Kooper and MacIntyre \cite{Kooper03}.} \label{fig:seven}
\end{figure}

\vspace{-5pt}
\section*{2001}

\begin{wrapfigure}{L}{2.1cm}
	\vspace{-10pt}	
	\includegraphics[width=1.0cm]{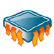}
	\includegraphics[width=1.0cm]{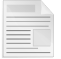}	
	\vspace{-20pt}		
\end{wrapfigure}
Joseph Newman \etal present the \textbf{BatPortal} \cite{Newman01}, a PDA-based, wireless AR system (see Fig.\ref{fig:sevena}). Localization is performed by measuring the travel time of ultra-sonic pulses between specially built devices worn by the user, so-called Bats, and fixed installed receivers deployed in the floors ceilings building-wide. The system can support an HMD-based system, but also the more well known BatPortal using a handheld device. Based on a fixed configuration of the PDA carried and the personal Bat worn, the direction of the users view is estimated, and a model of the scene with additional information about the scene is rendered onto the PDA screen.

\vspace{0.1in}

\begin{wrapfigure}{L}{2.1cm}
	\vspace{-10pt}	
	\includegraphics[width=1.0cm]{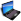}
	\includegraphics[width=1.0cm]{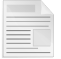}	
	\vspace{-20pt}		
\end{wrapfigure}
\noindent Hara \etal introduce \textbf{TOWNWEAR}, an outdoor system that uses a fiber optic gyroscope for orientation tracking \cite{Satoh01} (see Fig.\ref{fig:sevenb}). The high precision gyroscope is used to measure the 3DOF head direction accurately with minimal drift, which is then compensated by tracking natural features.

\vspace{0.1in}

\begin{wrapfigure}{L}{2.1cm}
	\vspace{-15pt}	
	\includegraphics[width=1.0cm]{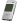}
	\includegraphics[width=1.0cm]{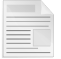}	
	\vspace{-25pt}		
\end{wrapfigure}
\noindent J\"urgen Fruend \etal present \textbf{AR-PDA}, a concept for building a wireless AR system and a special prototype of palm-sized hardware \cite{Fruend01} (see Fig.\ref{fig:sevenc}). Basic design ideas include the augmentation of real camera images with additional virtual objects, for example for illustration of functionality and interaction with commonly used household equipment.

\vspace{0.1in}

\begin{wrapfigure}{L}{2.1cm}
	\vspace{-10pt}	
	\includegraphics[width=1.0cm]{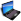}
	\includegraphics[width=1.0cm]{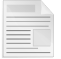}	
	\vspace{-20pt}		
\end{wrapfigure}
\noindent Reitmayr and Schmalstieg present \textbf{a mobile, multi-user AR system} \cite{Reitmayr01} (see Fig.\ref{fig:sevend}). The ideas of mobile augmented reality and collaboration between users in augmented shared space are combined and merged into a hybrid system. Communication is performed using LAN and wireless LAN, where mobile users and stationary users are acting in a common augmented space.

\vspace{0.1in}

\begin{wrapfigure}{L}{3.2cm}
	\vspace{-15pt}	
	\includegraphics[width=1.0cm]{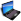}
	\includegraphics[width=1.0cm]{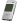}
	\includegraphics[width=1.0cm]{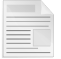}		
	\vspace{-25pt}		
\end{wrapfigure}
\noindent Vlahakis et al. present \textbf{Archeoguide}, a mobile AR system for cultural heritage sites \cite{Vlahakis2001} (see Fig.\ref{fig:sevene}). The system is built around the historical site of Olympia, Greece. The system contains a navigation interface, 3D models of ancient temples and statues, and avatars which are competing for the win in the historical run in the ancient Stadium. While communication is based on WLAN, accurate localization is performed using GPS. Within the system a scalable setup of mobile units can be used, starting with a notebook sized system with HMD, down to palmtop computers and Pocket PCs.

\vspace{0.1in}

\begin{wrapfigure}{L}{2.1cm}
	\vspace{-10pt}	
	\includegraphics[width=1.0cm]{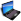}
	\includegraphics[width=1.0cm]{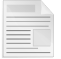}	
	\vspace{-20pt}		
\end{wrapfigure}
\noindent Kretschmer et al. present the \textbf{GEIST} system, a system for interactive story-telling within urban and/or historical environments \cite{Kretschmer01}. A complex database setup provides information queues for the appearance of buildings in ancient times or historical facts and events. Complex queries can be formulated and stories can be told by fictional avatars or historical persons.

\vspace{0.1in}

\begin{wrapfigure}{L}{2.1cm}
	\vspace{-10pt}	
	\includegraphics[width=1.0cm]{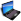}
	\includegraphics[width=1.0cm]{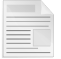}	
	\vspace{-20pt}		
\end{wrapfigure}
\noindent Columbia's Computer Graphics and User Interfaces Lab does an outdoor demonstration of their mobile AR restaurant guide at ISAR 2001, running on their Touring Machine \cite{Bell01} (see Fig.\ref{fig:sevenf}). Pop-up information sheets for nearby restaurants are overlaid on the user's view, and linked to reviews, menus, photos, and restaurant URLs.

\vspace{0.1in}

\begin{wrapfigure}{L}{2.1cm}
	\vspace{-10pt}	
	\includegraphics[width=1.0cm]{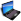}
	\includegraphics[width=1.0cm]{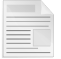}	
	\vspace{-20pt}		
\end{wrapfigure}
\noindent Kooper and MacIntyre create the \textbf{RWWW Browser}, a mobile AR application that acts as an interface to the World Wide Web \cite{Kooper03} (see Fig.\ref{fig:seveng}). It is the \textbf{first AR browser}. This early system suffers from the cumbersome AR hardware of that time, requiring a head mounted display and complicated tracking infrastructure. In 2008 Wikitude implements a similar idea on a mobile phone.

 \begin{figure}[tbp]
\centering
\vspace{-20pt}
\subfigure[]{\label{fig:eighta}\includegraphics[height=4.6cm]{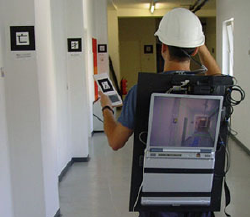}} \hfill
\subfigure[]{\label{fig:eightb}\includegraphics[height=4.6cm]{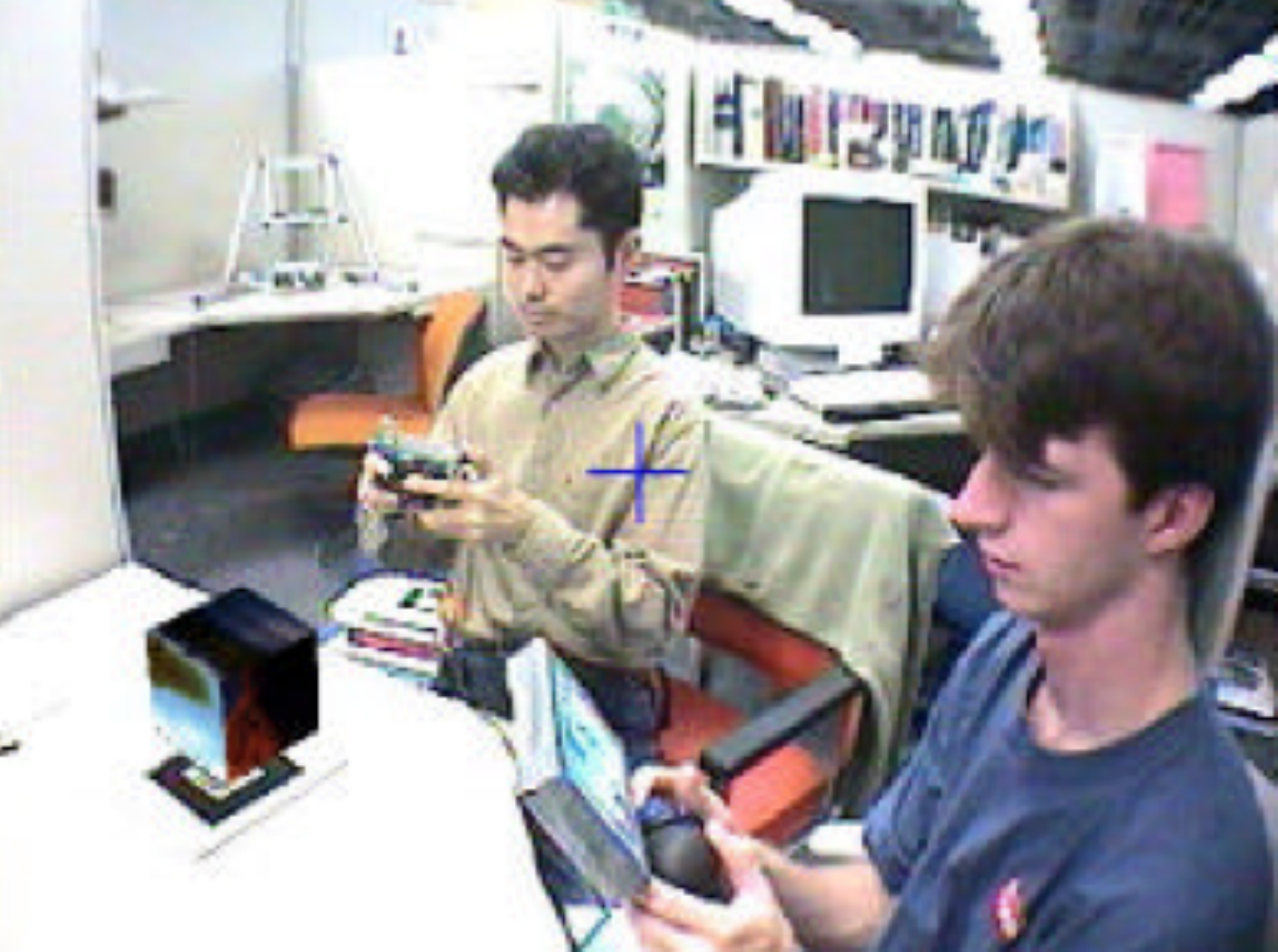}} \\
\subfigure[]{\label{fig:eightc}\includegraphics[height=4.6cm]{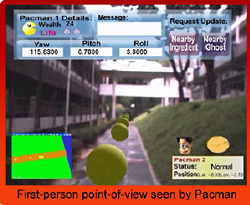}} \hfill
\subfigure[]{\label{fig:eightd}\includegraphics[height=4.6cm]{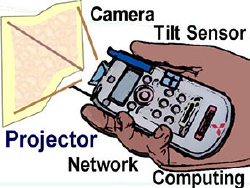}} \\
\subfigure[]{\label{fig:eighte}\includegraphics[height=3.7cm]{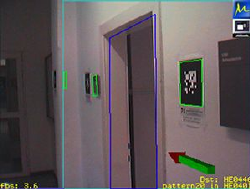}} \hfill
\subfigure[]{\label{fig:eightf}\includegraphics[height=3.7cm]{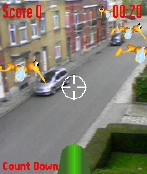}} \hfill
\subfigure[]{\label{fig:eightg}\includegraphics[height=3.7cm]{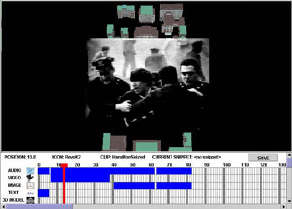}}
\vspace{-10pt}
\caption{(a): Navigation system by Kalkusch \etal \cite{Kalkusch02}. 
(b): ARPad by Mogilev \etal \cite{Mogilev02}. (c): Human Pacman by Cheok \etal \cite{Cheok03}. (d): iLamps system by Raskar \etal \cite{Raskar03}. (e): Indoor AR guidance system by Wagner and Schmalstieg \cite{Wagner03}. (f) Siemens SX1 AR game "Mozzies". (g): Mobile Authoring system by Guven and Feiner \cite{Guven03}.} \label{fig:eight}
\vspace{-0pt}
\end{figure}

\vspace{-5pt}
\section*{2002}

\begin{wrapfigure}{L}{2.1cm}
	\vspace{-10pt}	
	\includegraphics[width=1.0cm]{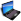}
	\includegraphics[width=1.0cm]{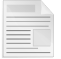}	
	\vspace{-20pt}		
\end{wrapfigure}
Michael Kalkusch \etal present a mobile augmented reality system to guide a user through an unfamiliar building to a destination room \cite{Kalkusch02} (see Fig. \ref{fig:eighta}). The system presents a world-registered wire frame model of the building labeled with directional information in a see-through heads-up display, and a three-dimensional world-in-miniature (WIM) map on a wrist-worn pad that also acts as an input device. Tracking is done using a combination of wall-mounted ARToolkit markers observed by a head-mounted camera, and an inertial tracker.

\vspace{0.1in}

\begin{wrapfigure}{L}{2.1cm}
	\vspace{-10pt}	
	\includegraphics[width=1.0cm]{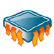}
	\includegraphics[width=1.0cm]{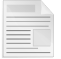}	
	\vspace{-20pt}		
\end{wrapfigure}
\noindent Leonid Naimark and Eric Foxlin present a \textbf{wearable low-power hybrid visual and inertial tracker} \cite{Naimark02}. This tracker, later to be known as InterSense’s IS-1200, can be used for tracking in large scale, such as a complete building. This is achieved by tracking a newly designed 2-D barcode with thousands of different codes and combining the result with an inertial sensor.

\vspace{0.1in}

\begin{wrapfigure}{L}{2.1cm}
	\vspace{-15pt}	
	\includegraphics[width=1.0cm]{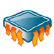}
	\includegraphics[width=1.0cm]{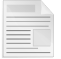}	
	\vspace{-30pt}		
\end{wrapfigure}
\noindent Mogilev \etal introduce the AR Pad, an ad-hoc mobile AR device equipped with a spaceball controller \cite{Mogilev02} (see Fig \ref{fig:eightb}).

\vspace{-5pt}
\section*{2003}

\begin{wrapfigure}{L}{2.1cm}
	\vspace{-10pt}	
	\includegraphics[width=1.0cm]{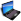}
	\includegraphics[width=1.0cm]{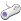}	
	\vspace{-20pt}		
\end{wrapfigure}
Adrian David Cheok \etal present the \textbf{Human Pacman} \cite{Cheok03} (see Fig. \ref{fig:eightc}). Human Pacman is an interactive ubiquitous and mobile entertainment system that is built upon position and perspective sensing via Global Positioning System and inertia sensors; and tangible human-computer interfacing with the use of Bluetooth and capacitive sensors. Pacmen and Ghosts are now real human players in the real world experiencing mixed computer graphics fantasy-reality provided by using wearable computers that are equipped with GPS and inertia sensors for players' position and perspective tracking. Virtual cookies and actual tangible physical objects with Bluetooth devices and capacitive sensors are incorporated into the game play to provide novel experiences of seamless transitions between real and virtual worlds.

\vspace{0.1in}

\begin{wrapfigure}{L}{2.1cm}
	\vspace{-10pt}	
	\includegraphics[width=1.0cm]{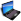}
	\includegraphics[width=1.0cm]{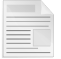}	
	\vspace{-20pt}		
\end{wrapfigure}
\noindent Ramesh Raskar \etal present \textbf{iLamps} \cite{Raskar03} (see Fig. \ref{fig:eightd}). This work created a first prototype for object augmentation with a hand-held projector-camera system. An enhanced projector can determine and respond to the geometry of the display surface, and can be used in an ad-hoc cluster to create a self-configuring display. Furthermore interaction techniques and co-operation between multiple units are discussed.

\vspace{0.1in}

\begin{wrapfigure}{L}{2.1cm}
	\vspace{-15pt}	
	\includegraphics[width=1.0cm]{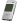}
	\includegraphics[width=1.0cm]{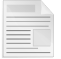}	
	\vspace{-25pt}		
\end{wrapfigure}
\noindent Daniel Wagner and Dieter Schmalstieg present an \textbf{indoor AR guidance system} running autonomously on a PDA \cite{Wagner03} (see Fig. \ref{fig:eighte}). They exploit the wide availability of consumer devices with a minimal need for infrastructure. The application provides the user with a three-dimensional augmented view of the environment by using a Windows Mobile port of ARToolKit for tracking and runs directly on the PDA.

\vspace{0.1in}

\begin{wrapfigure}{L}{2.1cm}
	\vspace{-0pt}	
	\includegraphics[width=1.0cm]{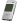}
	\includegraphics[width=1.0cm]{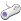}	
	\vspace{-30pt}		
\end{wrapfigure}
\noindent The Siemens SX1 is released, coming with the first commercial mobile phone AR camera game called \textbf{Mozzies} (also known as Mosquito Hunt) (see Fig. \ref{fig:eightf}). The mosquitoes are superimposed on the live video feed from the camera. Aiming is done by moving the phone around so that the cross hair points at the mosquitoes. Mozzies was awarded the title of best mobile game in 2003.
 
\vspace{0.1in}
 
\begin{wrapfigure}{L}{1.1cm}
	\vspace{-10pt}	
	\includegraphics[width=1.0cm]{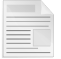}	
	\vspace{-10pt}		
\end{wrapfigure}
\noindent Sinem Guven presents a \textbf{mobile AR authoring system} for creating and editing 3D hypermedia narratives that are interwoven with a wearable computer user's surrounding environment\footnote{\url{http://graphics.cs.columbia.edu/projects/mars/Authoring.html}} \cite{Guven03} (see Fig. \ref{fig:eightg}). Their system was designed for authors who are not programmers and used a combination of 3D drag-and-drop for positioning media and a timeline for synchronization. It allowed authors to preview their results on a desktop workstation, as well as with a wearable AR or VR system.
 
\begin{figure}[tbp]
\centering
\vspace{-0pt}
\begin{tabular}{ccc}
\subfigure[]{\label{fig:ninea}\includegraphics[height=3.3cm]{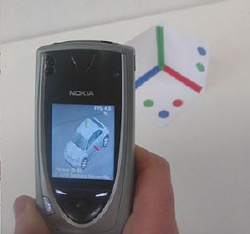}} &
\subfigure[]{\label{fig:nineb}\includegraphics[height=3.3cm]{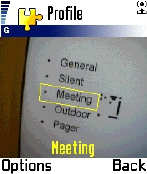}} & \multirow{-6}[1.0]{*}{
\subfigure[]{\label{fig:ninec}\includegraphics[height=8.0cm]{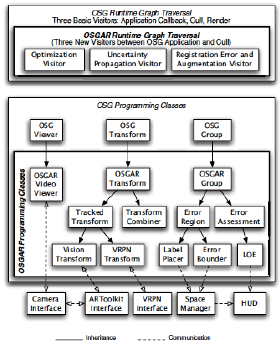}}} \\
\multicolumn{2}{c}{\subfigure[]{\label{fig:nined}\includegraphics[height=4.6cm]{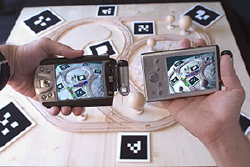}}} & \\
\end{tabular} 
\vspace{-10pt}
\caption{(a): Tracking 3D markers by M\"ohring \etal \cite{Mohring04}. (b): Visual Codes by Rohs and Gfeller \cite{Rohs04}. (c): OSGAR system by Coelho \etal \cite{Coelho04}. (d): The Invisible Train \cite{Wagner05}.} \label{fig:nine}
\end{figure}

\vspace{-5pt}
\section*{2004}

\begin{wrapfigure}{L}{2.1cm}
	\vspace{-15pt}	
	\includegraphics[width=1.0cm]{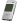}
	\includegraphics[width=1.0cm]{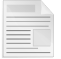}	
	\vspace{-25pt}		
\end{wrapfigure}
Mathias M\"ohring \etal present a system for \textbf{tracking 3D markers} on a mobile phone \cite{Mohring04} (see Fig.\ref{fig:ninea}). This work showed a first video see-through augmented reality system on a consumer cell-phone. It supports the detection and differentiation of different 3D markers, and correct integration of rendered 3D graphics into the live video stream.

\vspace{0.1in}

\begin{wrapfigure}{L}{2.1cm}
	\vspace{-15pt}	
	\includegraphics[width=1.0cm]{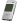}
	\includegraphics[width=1.0cm]{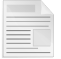}	
	\vspace{-25pt}		
\end{wrapfigure}
\noindent Michael Rohs and Beat Gfeller present \textbf{Visual Codes}, a 2D marker system for mobile phones \cite{Rohs04} (see Fig.\ref{fig:nineb}). These codes can be attached to physical objects in order to retrieve object-related information and functionality. They are also suitable for display on electronic screens.

\vspace{0.1in}

\begin{wrapfigure}{L}{1.1cm}
	\vspace{-10pt}	
	\includegraphics[width=1.0cm]{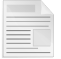}	
	\vspace{-10pt}		
\end{wrapfigure}
\noindent Enylton Machado Coelho \etal presents \textbf{OSGAR}, a scene graph with uncertain transformations \cite{Coelho04} (see Fig.\ref{fig:ninec}). In their work they target the problem of registration error, which is especially important for mobile scenarios when high quality tracking is not available and overlay graphics will not align perfectly with the real environment. OSGAR dynamically adapts the display to mitigate the effects of registration errors.

\vspace{0.1in} 

\begin{wrapfigure}{L}{2.1cm}
	\vspace{-10pt}	
	\includegraphics[width=1.0cm]{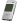}
	\includegraphics[width=1.0cm]{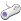}	
	\vspace{-20pt}		
\end{wrapfigure}
\noindent \textbf{The Invisible Train}, is shown at SIGGRAPH 2004 Emerging Technologies\footnote{The Invisible Train: \url{http://studierstube.icg.tugraz.at/invisible_train/}} (see Fig.\ref{fig:nined}). The Invisible Train is the first multi-user Augmented Reality application for handheld devices \cite{Wagner05}.

\vspace{-5pt}
\section*{2005}

\begin{wrapfigure}{L}{2.1cm}
	\vspace{-15pt}	
	\includegraphics[width=1.0cm]{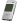}
	\includegraphics[width=1.0cm]{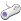}	
	\vspace{-25pt}		
\end{wrapfigure}
\begin{figure}[tbp]
\centering
\vspace{-0pt}
\subfigure[]{\label{fig:tena}\includegraphics[height=4.4cm]{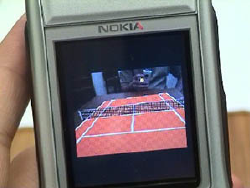}} \hfill
\subfigure[]{\label{fig:tenb}\includegraphics[height=4.4cm]{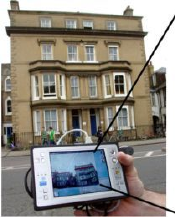}} \hfill 
\subfigure[]{\label{fig:tenc}\includegraphics[height=4.4cm]{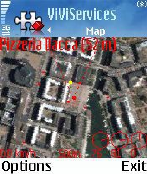}} \\
\vspace{-10pt}
\caption{(a): AR-Tennis by Henrysson \etal \cite{Henrysson05}. (b): Going Out by Reitmayr and Drummond \cite{Reitmayr06}. (c): Mara system by Nokia in 2006.} \label{fig:ten}
\end{figure}

Anders Henrysson ports ARToolKit to Symbian \cite{Henrysson05} (see Fig.\ref{fig:tena}). Based on this technology he presents the famous \textbf{AR-Tennis} game, the first collaborative AR application running on a mobile phone. ARTennis was awarded the Indepdent Mobile Gaming best game award for 2005, and the technical achievement award.

\vspace{0.1in}

\begin{wrapfigure}{L}{2.1cm}
	\vspace{-15pt}	
	\includegraphics[width=1.0cm]{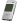}
	\includegraphics[width=1.0cm]{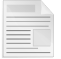}	
	\vspace{-25pt}		
\end{wrapfigure}
\noindent Project \textbf{ULTRA} shows how to use non-realtime natural feature tracking on PDAs to support people in multiple domains such as the maintenance and support of complex machines, construction and production, and edutainment and cultural heritage \cite{Makri05}. Furthermore an authoring environment is developed to create the AR scenes for the maintenance tasks.

\vspace{0.1in}

\begin{wrapfigure}{L}{1.1cm}
	\vspace{-0pt}	
	\includegraphics[width=1.0cm]{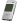}
	\vspace{-20pt}		
\end{wrapfigure}
\noindent The first \textbf{mobile phones equipped with three-axis accelerometers} were the Sharp V603SH and the Samsung SCH-S310 both sold in Asia in 2005. 

\vspace{-5pt} 
\section*{2006}

\begin{wrapfigure}{L}{2.1cm}
	\vspace{-10pt}	
	\includegraphics[width=1.0cm]{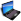}
	\includegraphics[width=1.0cm]{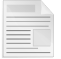}	
	\vspace{-20pt}		
\end{wrapfigure}
Reitmayr and Drummond present a model-based hybrid \textbf{tracking system for outdoor augmented reality} in urban environments enabling accurate, real-time overlays on a handheld device \cite{Reitmayr06} (see Fig.\ref{fig:tenb}). The system combines an edge-based tracker for accurate localization, gyroscope measurements to deal with fast motions, measurements of gravity and magnetic field to avoid drift, and a back store of reference frames with online frame selection to re-initialize automatically after dynamic occlusions or failures.

\vspace{0.1in} 

\begin{wrapfigure}{L}{2.1cm}
	\vspace{-15pt}	
	\includegraphics[width=1.0cm]{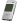}
	\includegraphics[width=1.0cm]{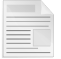}	
	\vspace{-25pt}		
\end{wrapfigure}
\noindent Nokia presents Mara, a \textbf{multi-sensor mobile phone AR} guidance application for mobile phones\footnote{Mara: \url{http://research.nokia.com/page/219}}. The prototype application overlays the continuous viewfinder image stream captured by the camera with graphics and text in real time, annotating the user's surroundings (see Fig.\ref{fig:tenc}).

\newpage

\vspace{-5pt} 
\section*{2007}

\begin{wrapfigure}{L}{2.1cm}
	\vspace{-10pt}	
	\includegraphics[width=1.0cm]{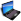}
	\includegraphics[width=1.0cm]{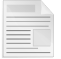}	
	\vspace{-20pt}		
\end{wrapfigure}
\begin{figure}[tbp]
\centering
\vspace{-70pt}
\subfigure[]{\label{fig:elevena}\includegraphics[height=4.3cm]{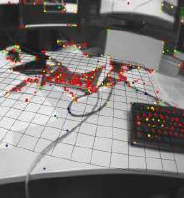}} \hfill
\subfigure[]{\label{fig:elevenb}\includegraphics[height=4.3cm]{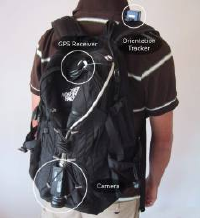}} \hfill 
\subfigure[]{\label{fig:elevenc}\includegraphics[height=4.3cm]{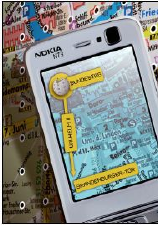}} \\ \hfill
\subfigure[]{\label{fig:elevend}\includegraphics[trim=1cm 1cm 1cm 1cm, clip=true,height=4.3cm]{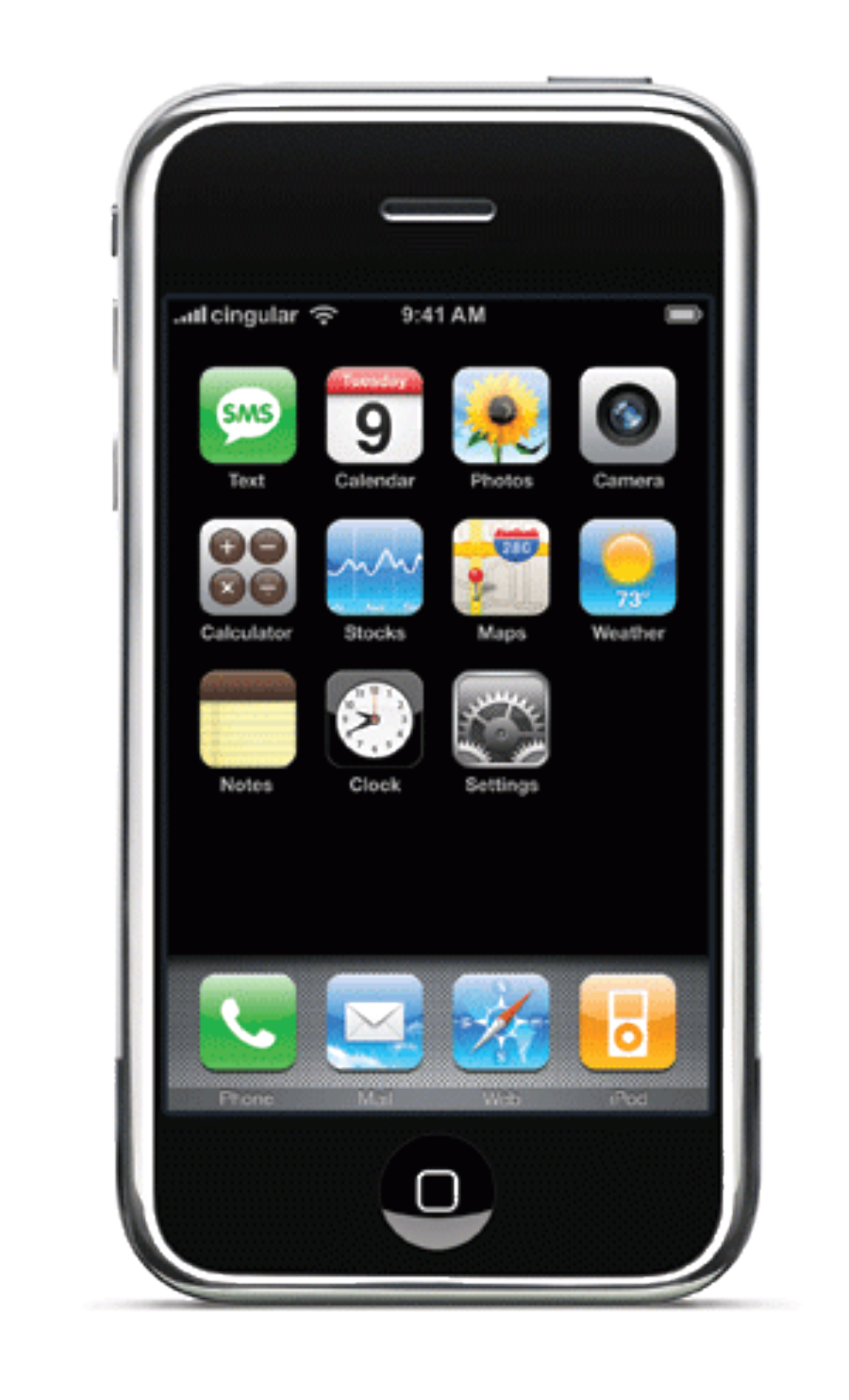}} \hfill
\subfigure[]{\label{fig:elevene}\includegraphics[height=4.3cm]{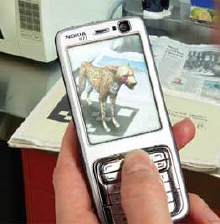}} \hfill
\vspace{-10pt}
\caption{(a): PTAM by Klein and Murray \cite{Klein07}. (b): Groundcam by DiVerdi and H\"ollerer \cite{DiVerdi07}. (c): Map Navigation with mobile devices by Rohs \etal \cite{Rohs07}. (d): Apple iPhone 2G. (e): AR advertising app by HIT Lab NZ and Saatchi.} \label{fig:eleven}
\end{figure}

Klein and Murray present a system capable of \textbf{robust real-time tracking and mapping} in parallel with a monocular camera in small workspaces \cite{Klein07} (see Fig. \ref{fig:elevena}). It is an adaption of a SLAM approach which processes the tracking and mapping task on two separated threads.

\vspace{0.1in}

\begin{wrapfigure}{L}{2.1cm}
	\vspace{-10pt}	
	\includegraphics[width=1.0cm]{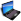}
	\includegraphics[width=1.0cm]{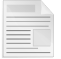}	
	\vspace{-20pt}		
\end{wrapfigure}
\noindent DiVerdi and H\"ollerer present the \textbf{GroundCam}, a system combining a camera and an orientation tracker \cite{DiVerdi07} (see Fig. \ref{fig:elevenb}). The camera points at the ground behind the user and provides 2D tracking information. The method is similar to that of an optical desktop mouse. 

\vspace{0.1in}

\begin{wrapfigure}{L}{2.1cm}
	\vspace{-15pt}	
	\includegraphics[width=1.0cm]{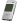}
	\includegraphics[width=1.0cm]{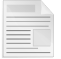}	
	\vspace{-25pt}		
\end{wrapfigure}
\noindent Rohs \etal compare the performance of the following navigation methods for map navigation on mobile devices: joystick navigation, the dynamic peephole method without visual context, and the magic lens paradigm using external visual context \cite{Rohs07} (see Fig. \ref{fig:elevenc}). In their user study they demonstrate the advantage of dynamic peephole and magic lens interaction over joystick interaction in terms of search time and degree of exploration of the search space.

\vspace{0.1in}

\begin{wrapfigure}{L}{1.1cm}
	\vspace{-0pt}	
	\includegraphics[width=1.0cm]{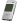}
	\vspace{-20pt}		
\end{wrapfigure}
\noindent The \textbf{first multi-touch screen mobile phone}, famously known as iPhone sold by Apple, leverages a new way to interact on mobile devices (see Fig. \ref{fig:elevend}).

\vspace{0.1in}

\begin{wrapfigure}{L}{1.1cm}
	\vspace{-15pt}	
	\includegraphics[width=1.0cm]{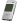}
	\vspace{-20pt}		
\end{wrapfigure}
\noindent HIT Lab NZ and Saatchi deliver the world's \textbf{first mobile phone based AR advertising} application for the Wellington Zoo \citelinks{HIT08} (see Fig. \ref{fig:elevene}).

\begin{figure}[tbp]
\centering
\vspace{-50pt}
\subfigure[]{\label{fig:twelvea}\includegraphics[height=3.5cm]{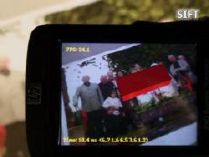}} \hfill
\subfigure[]{\label{fig:twelveb}\includegraphics[height=3.5cm]{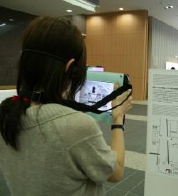}} \hfill 
\subfigure[]{\label{fig:twelvec}\includegraphics[height=3.5cm]{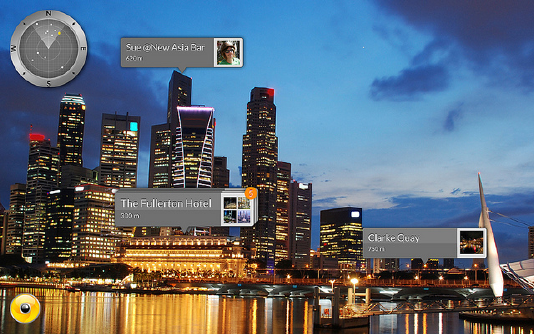}}
\vspace{-10pt}
\caption{(a): Real-time natural feature tracking on mobile phones by Wagner \etal \cite{Wagner08}. (b): Commercial AR museum guide by METAIO \cite{Miyashita08}.  (c): Wikitude AR Browser.} \label{fig:twelve}
\end{figure}

\vspace{-5pt}
\section*{2008}

\begin{wrapfigure}{L}{2.1cm}
	\vspace{-15pt}	
	\includegraphics[width=1.0cm]{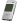}
	\includegraphics[width=1.0cm]{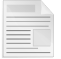}	
	\vspace{-25pt}		
\end{wrapfigure}
Wagner \etal present the \textbf{first 6DOF implementation of natural feature tracking in real-time on mobile phones} achieving interactive frame rates of up to 20 Hz \cite{Wagner08} (see Fig. \ref{fig:twelvea}). They heavily modify the well known SIFT and Ferns methods in order to gain more speed and reduce memory requirements.

\vspace{0.1in}

\begin{wrapfigure}{L}{2.1cm}
	\vspace{-10pt}	
	\includegraphics[width=1.0cm]{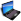}
	\includegraphics[width=1.0cm]{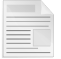}	
	\vspace{-20pt}		
\end{wrapfigure}
\noindent METAIO presents a \textbf{commercial mobile AR museum guide} using natural feature tracking or a six-month exhibition on Islamic art \cite{Miyashita08} (see Fig. \ref{fig:twelveb}). In their paper they describe the experiences made in this project.

\vspace{0.1in}

\begin{wrapfigure}{L}{1.1cm}
	\vspace{-10pt}	
	\includegraphics[width=1.0cm]{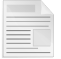}	
	\vspace{-10pt}		
\end{wrapfigure}
\noindent With Augmented Reality 2.0, Schmalstieg \etal presented at the Dagstuhl seminar in 2008 for the first time a concept that combined ideas of the Web 2.0 such as social media, crowd sourcing through public participation, and an open architecture for content markup and distribution, and applied it to mobile Augmented Reality to create a scalable AR experience \cite{Schmalstieg11}.

\newpage

\begin{wrapfigure}{L}{2.1cm}
	\vspace{-0pt}	
	\includegraphics[width=1.0cm]{figs/icons/phone}
	\includegraphics[width=1.0cm]{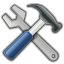}	
	\vspace{-20pt}		
\end{wrapfigure}
\noindent Mobilizy launches \textbf{Wikitude}\footnote{Wikitude: \url{http://www.mobilizy.com/wikitude.php?lang=en}}, an application that combines GPS and compass data with Wikipedia entries. The Wikitude World Browser overlays information on the real-time camera view of an Android smartphone (see Fig. \ref{fig:twelvec}).

\vspace{-5pt} 
\section*{2009}

\begin{wrapfigure}{L}{2.1cm}
	\vspace{-15pt}	
	\includegraphics[width=1.0cm]{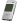}
	\includegraphics[width=1.0cm]{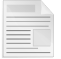}	
	\vspace{-25pt}		
\end{wrapfigure}
\begin{figure}[tbp]
\centering
\vspace{-35pt}
\subfigure[]{\label{fig:thirteena}\includegraphics[height=5.1cm]{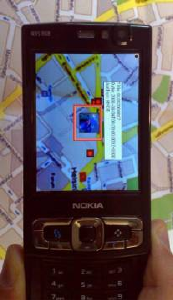}} \hfill
\subfigure[]{\label{fig:thirteenb}\includegraphics[height=5.1cm]{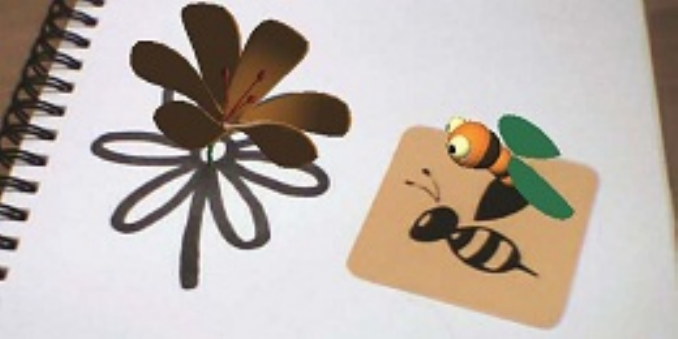}}\\  \hfill
\subfigure[]{\label{fig:thirteenc}\includegraphics[height=5.1cm]{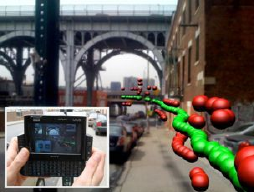}} \hfill 
\subfigure[]{\label{fig:thirteend}\includegraphics[height=5.1cm]{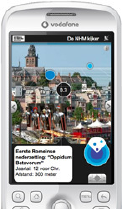}}  \\
\subfigure[]{\label{fig:thirteene}\includegraphics[height=5.1cm]{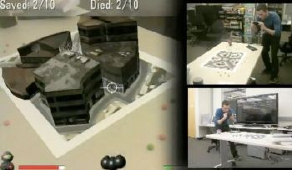}} \hfill 
\subfigure[]{\label{fig:thirteenf}\includegraphics[height=5.1cm]{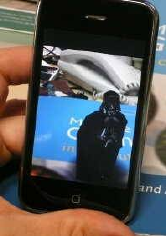}}
\vspace{-10pt}
\caption{(a): MapLens by Morrison \etal \cite{Morrison09}. (b) Hagbi's pose tracking using shape \cite{Bergig09}. (c): SiteLens by White and Feiner \cite{White09}. (d): LayAR AR browser. (e): ARhrrrr! Zombie game by Spreen \etal from Georgia Tech. (f): Klein's PTAM system running on an iPhone \cite{Klein09}. } \label{fig:thirteen}
\end{figure}

Morrison \etal present MapLens which is a mobile augmented reality (AR) map using a \textbf{magic lens over a paper map} \cite{Morrison09} (see Fig. \ref{fig:thirteena}). They conduct a broad user study in form of an outdoor location-based game. Their main finding is that AR features facilitate place-making by creating a constant need for referencing to the physical. The field trials show that the main potential of AR maps lies in their use as a collaborative tool.

\vspace{0.1in}

\begin{wrapfigure}{L}{2.1cm}
	\vspace{-12pt}	
	\includegraphics[width=1.0cm]{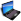}
	\includegraphics[width=1.0cm]{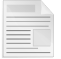}	
	\vspace{-28pt}		
\end{wrapfigure}
\noindent Hagbi \etal presented an approach allowing to track the pose of the mobile device by pointing it to fiducials \cite{Bergig09} (see Fig. \ref{fig:thirteenb}). Unlike existing systems the approach allows to track a wide set of planar shapes while the user can teach the system new shapes at runtime by showing them to the camera. The learned shapes are then maintained by the system in a shape library enabling new AR application scenarios in terms of interaction with the scene but also in terms of fiducial design.

\vspace{0.1in}

\begin{wrapfigure}{L}{2.1cm}
	\vspace{-15pt}	
	\includegraphics[width=1.0cm]{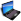}
	\includegraphics[width=1.0cm]{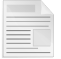}	
	\vspace{-25pt}		
\end{wrapfigure}
\noindent Sean White introduces SiteLens (see Fig. \ref{fig:thirteenc}), a \textbf{hand-held mobile AR system for urban design and urban planning site visits} \cite{White09}. SiteLens creates "situated visualizations" that are related to and displayed in their environment. For example, representations of geocoded carbon monoxide concentration data are overlaid at the sites at which the data
was recorded.

\vspace{0.1in}

\begin{wrapfigure}{L}{2.1cm}
	\vspace{-15pt}	
	\includegraphics[width=1.0cm]{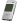}
	\includegraphics[width=1.0cm]{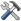}	
	\vspace{-25pt}		
\end{wrapfigure}
\noindent SPRXmobile launches Layar\footnote{LayAR: \url{http://layar.eu/}}, an \textbf{advanced variant} of Wikitude (see Fig. \ref{fig:thirteend}). Layar uses the same registration mechanism as Wikitude (GPS + compass), and incoperates this into an open client-server platform. Content layers are the equivalent of web pages in normal browsers. Existing layers include Wikipedia, Twitter and Brightkite to local services like Yelp, Trulia, store locators, nearby bus stops, mobile coupons, Mazda dealers and tourist, nature and cultural guides. On August 17th Layar went global serving almost 100 content layers.

\vspace{0.1in}

\begin{wrapfigure}{L}{2.1cm}
	\vspace{-0pt}	
	\includegraphics[width=1.0cm]{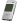}
	\includegraphics[width=1.0cm]{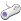}	
	\vspace{-30pt}		
\end{wrapfigure}
\noindent Kimberly Spreen \etal develop ARhrrrr!, the \textbf{first mobile AR game with high quality content} at the level of commercial games\citelinks{GT09} (see Fig. \ref{fig:thirteene}). They use an NVIDIA Tegra developer kit ("Concorde") with a fast GPU. All processing except for tracking are running on the GPU, making the whole application run at high frame rates on a mobile phone class device despite the highly detailed content and natural feature tracking.

\vspace{0.1in}

\begin{wrapfigure}{L}{2.1cm}
	\vspace{-15pt}	
	\includegraphics[width=1.0cm]{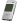}
	\includegraphics[width=1.0cm]{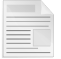}	
	\vspace{-25pt}		
\end{wrapfigure}
\noindent Georg Klein presents a video showing his \textbf{SLAM system running in real-time on an iPhone} \cite{Klein09} (see Fig. \ref{fig:thirteenf}) and later presents this at ISMAR 2009 in Orlando, Florida. Even though it has constrains in terms of working area it is the first time a 6DoF SLAM system is known to run on mobile phones in sufficient speed.

\vspace{0.2in}

\noindent \fcolorbox{black}{lightgray}{\begin{minipage}{\textwidth}
\paragraph{Update April 2015:} The following parts of the document until beginning of 2015 cover the years since the last homepage update, following the same categorization and scheme as before. 

From end of 2009 onwards, AR research and development is generally driven by high expectations and huge investments from world-leading companies such as Microsoft, Google, Facebook, Qualcomm and others. At the same time, the landscape of mobile phone manufacturers started to change radically.

In general the advances in mobile device capabilities introduce a strong drive towards mobile computing, and the availability of cloud processing further supports the proposal and development of server-client solutions for AR purposes. One major trend starting around 2010, originating by the work of Davison in 2003 \cite{Davison03} and later further explored by Klein and Murray \cite{Klein07,Klein09}, is the heavy use of SLAM in AR, which still continues to dominate a major part of AR research and development as of beginning of 2015.
\end{minipage}}

\vspace{0.2in}

\begin{wrapfigure}{L}{1.1cm}
	\vspace{-15pt}	
	\includegraphics[width=1.0cm]{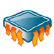}
	\vspace{-15pt}		
\end{wrapfigure}
\noindent Microsoft presents "Project Natal" at the game exhibition E3. It is the first version of a \textbf{new hardware interface}, consisting of \textbf{motion detection technology}, microphone, color camera and software, to be integrated into the game console Xbox 360.

\vspace{0.1in}

\begin{wrapfigure}{L}{2.1cm}
	\vspace{-15pt}	
	\includegraphics[width=1.0cm]{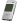}
	\includegraphics[width=1.0cm]{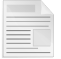}	
	\vspace{-25pt}		
\end{wrapfigure}
\begin{figure}[tbp]
\centering
\vspace{-25pt}
\subfigure[]{\label{fig:fourteena}\includegraphics[height=5.6cm]{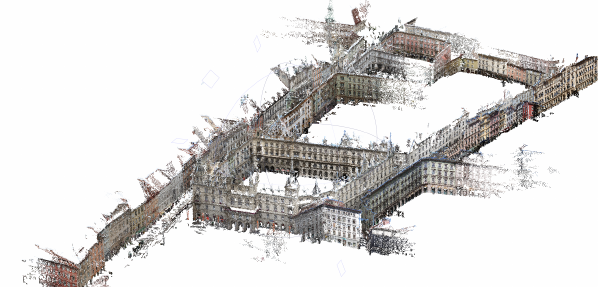}} \\\
\subfigure[]{\label{fig:fourteenb}\includegraphics[height=3.2cm]{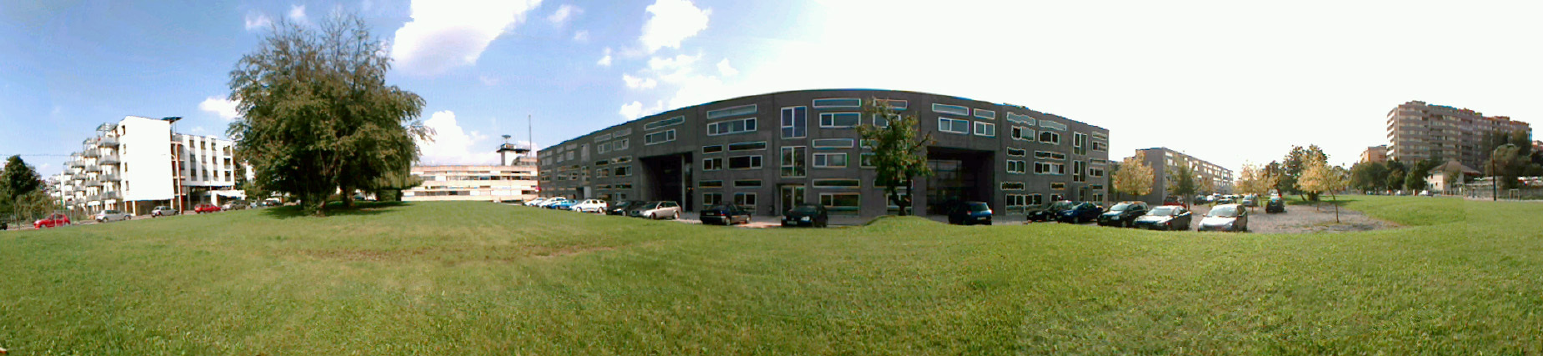}} \\
\subfigure[]{\label{fig:fourteenc}\includegraphics[height=4.6cm]{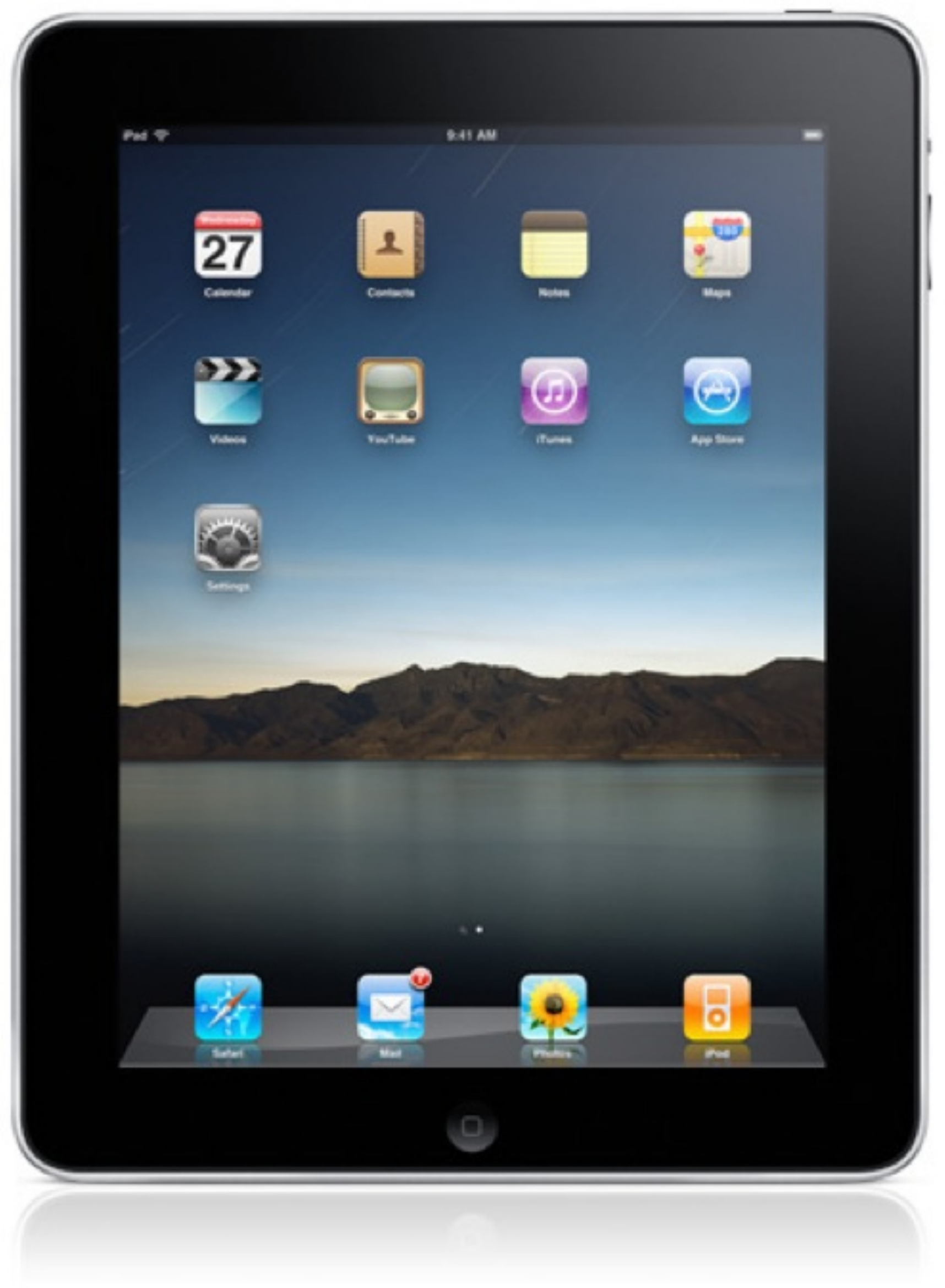}} \hfill
\subfigure[]{\label{fig:fourteend}\includegraphics[height=4.6cm]{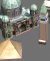}}  \hfill
\subfigure[]{\label{fig:fourteene}\includegraphics[height=4.6cm]{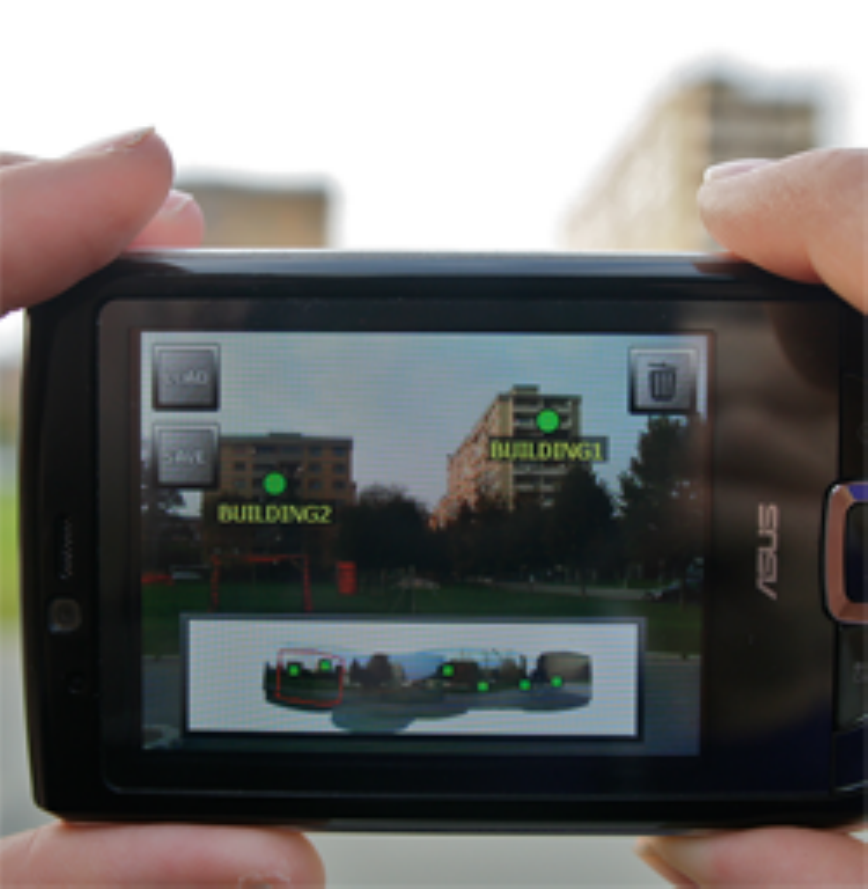}} \\
\vspace{-10pt}
\caption{(a): City reconstruction as used by Arth \etal \cite{Arth09}. (b): Panoramic image captured on a mobile phone using the approach of Wagner \etal \cite{Wagner10}. (c) Apple iPad. (d): City-of-Sights paperboard models by Gruber \etal \cite{Gruber10}. (e) In-situ information creation by Langlotz \etal \cite{Langlotz12}.} \label{fig:fourteen}
\vspace{-10pt}
\end{figure}

 
\noindent At ISMAR 2009, Clemens Arth \etal present a system for \textbf{large-scale localization and subsequent 6DOF tracking} on mobile phones \cite{Arth09}. The system uses sparse point clouds of city areas and FAST corners and SURF-like descriptors that can be used on memory-limited devices (see Fig. \ref{fig:fourteena}).

\newpage

\begin{wrapfigure}{L}{1.2cm}
	\vspace{0pt}	
	\includegraphics[width=1.2cm]{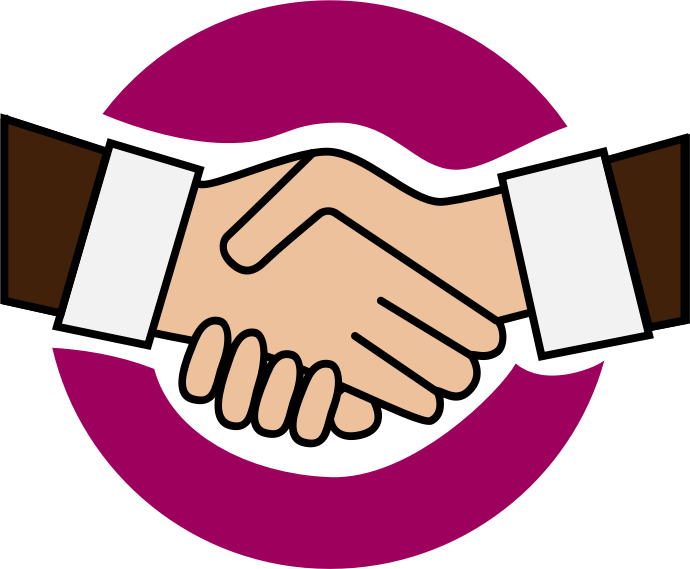}
	\vspace{-20pt}		
\end{wrapfigure} 
\noindent \textbf{Qualcomm Inc. acquires the mobile AR IP} from Imagination Computer Services GmbH., Vienna, and takes over the funding of the Christian Doppler Laboratory for Handheld AR at Graz University of Technology. A research center to focus on AR is opened later in 2010 in Vienna \citelinks{QC10}.

\vspace{-5pt}
\section*{2010}

\begin{wrapfigure}{L}{2.1cm}
	\vspace{-15pt}	
	\includegraphics[width=1.0cm]{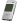}
	\includegraphics[width=1.0cm]{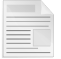}
	\vspace{-25pt}		
\end{wrapfigure} 
A \textbf{real-time panoramic mapping and tracking system for mobile phones} is presented by Wagner \etal at VR, which performs 3DOF tracking in cylindric space and supports the use of panoramic imagery for improved usability and experience in AR \cite{Wagner10} (see Fig. \ref{fig:fourteenb}).

\vspace{0.1in}

\begin{wrapfigure}{L}{2.1cm}
	\vspace{-15pt}	
	\includegraphics[width=1.0cm]{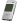}
	\includegraphics[width=1.0cm]{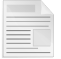}
	\vspace{-25pt}		
\end{wrapfigure} 
\noindent KHARMA is a lightweight and open architecture for referencing and delivering content explicitly aiming for mobile AR applications running on a global scale. It uses KML for describing the geospatial or relative relation of content while utilizing on HTML, JavaScript and CSS technologies for content development and delivery \cite{Hill10}.

\vspace{0.1in}

\begin{wrapfigure}{L}{2.3cm}
	\vspace{-15pt}	
	\includegraphics[width=1.2cm]{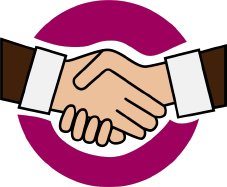}
	\includegraphics[width=1.0cm]{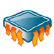}
	\vspace{-31pt}		
\end{wrapfigure} 
\noindent Microsoft announces a close cooperation with Primesense \citelinks{MS10}, an Israeli company working on structured-light based 3D sensors, to supply their technology to "Project Natal", now coined Kinect. The \textbf{Kinect becomes commercially available} in November 2010.

\vspace{0.1in}

\begin{wrapfigure}{L}{1.1cm}
	\vspace{-15pt}	
	\includegraphics[width=1.0cm]{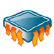}
	\vspace{-15pt}		
\end{wrapfigure} 
\noindent Apple releases the iPad\footnote{Wikipedia: \url{http://en.wikipedia.org/wiki/IPad}} on April 2010, which becomes the \textbf{first tablet computer to be adopted by the large public}. The iPad featured an assisted GPS, accelerometers, magnetometers, advanced graphics chipset (PowerVR SGX535), enabling the possibilities to create efficient AR application on tablet computer (see Fig. \ref{fig:fourteenc}).

\vspace{0.1in}

\begin{wrapfigure}{L}{1.1cm}
	\vspace{-10pt}	
	\includegraphics[width=1.0cm]{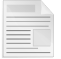}
	\vspace{-0pt}		
\end{wrapfigure} 
\noindent At ISMAR Lukas Gruber \etal present the \textbf{"City of Sights"}, a collection of datasets and paperboard models\footnote{\url{http://studierstube.icg.tugraz.at/handheld_ar/cityofsights.php}} to evaluate the tracking and reconstruction performance of algorithms used in AR \cite{Gruber10} (see Fig. \ref{fig:fourteend}).

\vspace{0.1in}

\begin{wrapfigure}{L}{1.1cm}
	\vspace{-15pt}	
	\includegraphics[width=1.0cm]{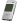}
	\vspace{-20pt}		
\end{wrapfigure}
\noindent After several delays, \textbf{Microsoft releases Windows Phone} in October 2010, to become the third major mobile phone operating system to challenge iOS and Android. 
\newpage

\begin{wrapfigure}{L}{2.1cm}
	\vspace{-0pt}	
	\includegraphics[width=1.0cm]{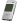}
	\includegraphics[width=1.0cm]{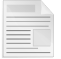}
	\vspace{-25pt}		
\end{wrapfigure} 
\noindent Existing mobile AR applications where exclusively used to browser and consume digital information. Langlotz et al. presented an new approach aiming for AR browsers that also supported creation of digital information in-situ. The information is registered with pixel-precision by utilizing a panorama of the environment that is created in the background \cite{Langlotz12} (see Fig. \ref{fig:fourteene}).

\vspace{-5pt}
\section*{2011}

\begin{wrapfigure}{L}{2.2cm}
	\vspace{-15pt}	
	\includegraphics[width=1.0cm]{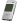}
	\includegraphics[width=1.0cm]{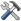}
	\vspace{-20pt}		
\end{wrapfigure} 
Qualcomm announces the \textbf{release of its AR platform SDK} to the public in April. At that time it is called QCAR \citelinks{QC11}, which will later be called Vuforia.

\vspace{0.1in}

\begin{wrapfigure}{L}{1.2cm}
	\vspace{-10pt}	
	\includegraphics[width=1.2cm]{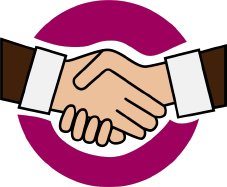}
	\vspace{-10pt}		
\end{wrapfigure} 
\noindent In August, \textbf{Google announces the acquisition of Motorola Mobility} for about \$12.5 billion \citelinks{GO11}. A major asset of Motorola is a large patent portfolio, which Google needs to secure the further Android platform development. 

\vspace{0.1in}

\begin{wrapfigure}{L}{2.1cm}
	\vspace{-10pt}	
	\includegraphics[width=1.0cm]{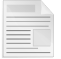}
	\includegraphics[width=1.0cm]{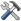}
	\vspace{-20pt}		
\end{wrapfigure}
\begin{figure}[tbp]
\centering
\vspace{-10pt}
\subfigure[]{\label{fig:fifteena}\includegraphics[height=4.7cm]{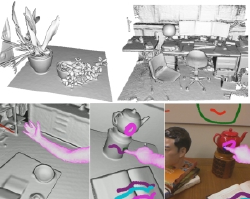}} \hfill 
\subfigure[]{\label{fig:fifteenb}\includegraphics[height=4.7cm]{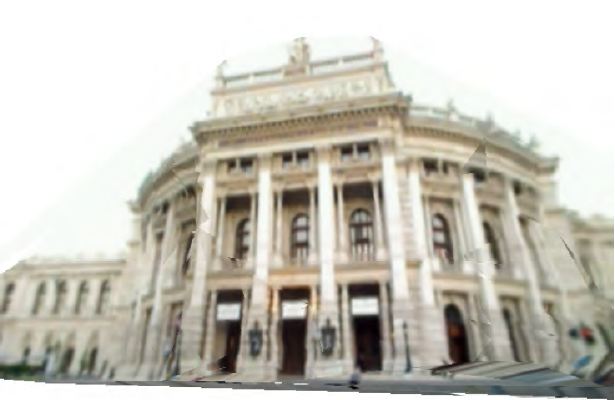}}
\vspace{-10pt}
\caption{ (a): KinectFusion system presented by Newcombe \etal at ISMAR 2011 \cite{Newcombe11b}. (b): Mobile phone scene reconstruction by Pan \etal \cite{Pan11}.}
\label{fig:fifteen}
\end{figure}

\noindent At ICCV 2011, Newcombe presents DTAM, a \textbf{dense real-time tracking and mapping algorithm} \cite{Newcombe11a}. Later at ISMAR 2011, Richard Newcombe presents the \textbf{KinectFusion} work \cite{Newcombe11b}, in which depth images from the Kinect sensor are fused to create a single implicit surface model. KinectFusion becomes publicly available within the Kinect SDK later \citelinks{KF11} (see Fig. \ref{fig:fifteena}).

\vspace{0.1in}

\begin{wrapfigure}{L}{2.1cm}
	\vspace{-15pt}	
	\includegraphics[width=1.0cm]{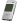}	
	\includegraphics[width=1.0cm]{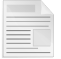}
	\vspace{-30pt}		
\end{wrapfigure}
\noindent Qi Pan presents his work on \textbf{reconstructing scenes on mobile phones} using panoramic images. By using FAST corners and a SURF-like descriptor, multiple panoramas are registered and a triangulated model is created after voxel carving \cite{Pan11} (see Fig. \ref{fig:fifteenb}).

\vspace{0.1in}

\begin{wrapfigure}{L}{2.1cm}
	\vspace{-15pt}	
	\includegraphics[width=1.0cm]{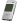}	
	\includegraphics[width=1.0cm]{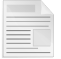}
	\vspace{-25pt}		
\end{wrapfigure}
\noindent Following the still challenging problem of running SLAM in real-time on mobiles, Pirchheim presents an approach using \textbf{planarity assumptions}, and demonstrates his approach on a Nokia N900 smartphone \cite{Pirchheim11}.

\vspace{0.1in}

\begin{wrapfigure}{L}{2.1cm}
	\vspace{-15pt}	
	\includegraphics[width=1.0cm]{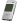}	
	\includegraphics[width=1.0cm]{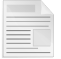}
	\vspace{-25pt}		
\end{wrapfigure}
\noindent Grubert \etal publish a technical report about the plausibility of using AR browsers \cite{Grubert11}, which becomes a survey about the pros and cons of AR browser technology at that point in time.

\vspace{-0pt}
\section*{2012}

\begin{wrapfigure}{L}{1.1cm}
	\vspace{-15pt}	
	\includegraphics[width=1.0cm]{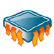}
	\vspace{-15pt}		
\end{wrapfigure} 
\textbf{Smart watches} are broadly introduced as a new generation of mobile wearables. \emph{Pebble} and the \emph{Sony SmartWatch} are built to connect to a personal smartphone and to provide simple functionality, such as notifications or call answering.

\vspace{0.1in}

\begin{wrapfigure}{L}{1.1cm}
	\vspace{-15pt}	
	\includegraphics[width=1.0cm]{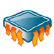}
	\vspace{-15pt}		
\end{wrapfigure} 
\noindent \textbf{Google Glass} (also known as Google \emph{Project Glass}) is firstly presented to the public\footnote{Google Glass project page on Google+: \url{https://plus.google.com/+GoogleGlass}} (see Fig.\ref{fig:sixteenb}). Goggle Glass is is an optical HMD that can be controlled with an integrated touch-sensitive sensor or natural language commands. After it's public announcement Google Glass had a major impact on research but even more on the public perception of mixed reality technology. 

\vspace{0.1in}

\begin{wrapfigure}{L}{1.1cm}
	\vspace{-15pt}	
	\includegraphics[width=1.0cm]{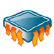}
	\vspace{-15pt}		
\end{wrapfigure} 
\noindent NVidia is demonstrating at Siggraph Emerging Technologies their prototype of a head mounted display supporting accurate accommodation, convergence, and binocular-disparity depth cues (see Fig. \ref{fig:sixteenc}). The prototype introduces a light-field-based approach to near-eye displays and can be seen as a next generation wearable display technology for AR as existing hardware can't provide accurate acommodation \citelinks{NV12}.

\vspace{0.1in}

\begin{wrapfigure}{L}{1.1cm}
	\vspace{-15pt}	
	\includegraphics[width=1.0cm]{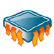}
	\vspace{-15pt}		
\end{wrapfigure} 
\noindent 13th lab released the first commercial mobile SLAM (Simultaneous localization and mapping) system coined Pointcloud\footnote{Pointcloud homepage: \url{http://pointcloud.io/}} to the public, marking a major milestone for app developers who want to integrate SLAM-based tracking into their application\footnote{Pointcloud video: \url{http://www.youtube.com/watch?v=K5OKaK3Ay8U}}.

\vspace{0.1in}

\begin{wrapfigure}{L}{1.1cm}
	\vspace{-15pt}	
	\includegraphics[width=1.0cm]{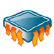}
	\vspace{-15pt}		
\end{wrapfigure} 
\noindent PrimeSense, the creator of the Microsoft Kinect, introduced a smaller version of a 3D sensing device called Capri \citelinks{PS12} that is small enough to be integrated into mobile devices such as tablets or smartphones\footnote{Capri Video: \url{http://www.youtube.com/watch?v=ELTETXO02zE}}.

\vspace{0.1in}

\begin{wrapfigure}{L}{2.1cm}
	\vspace{-15pt}	
	\includegraphics[width=1.0cm]{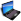}	
	\includegraphics[width=1.0cm]{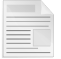}
	\vspace{-25pt}		
\end{wrapfigure}
\noindent At ISMAR 2012, Steffen Gauglitz \etal present their approach on \textbf{tracking and mapping from both general and rotation-only camera motion} \cite{Gauglitz12}. 

\vspace{0.1in}

\begin{wrapfigure}{L}{1.1cm}
	\vspace{-15pt}	
	\includegraphics[width=1.0cm]{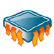}
	\vspace{-15pt}		
\end{wrapfigure} 
\noindent In August, Oculus VR announces the \textbf{Oculus Rift} dev kit, a virtual reality head-mounted display. This initiated a new hype in Virtual Reality and in the development of more head-mounted displays for gaming purposes mainly (see Fig.\ref{fig:sixteena}).

\vspace{-0pt}
\section*{2013}

\begin{wrapfigure}{L}{2.1cm}
	\vspace{-15pt}	
	\includegraphics[width=1.0cm]{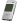}	
	\includegraphics[width=1.0cm]{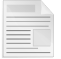}
	\vspace{-25pt}		
\end{wrapfigure}
As opposed to previous work from Gauglitz \etal, Pirchheim \etal present an approach to \textbf{handle pure camera rotation} running on \textbf{a mobile phone} at ISMAR \cite{Pirchheim13}.

\vspace{0.1in}

\begin{figure}[tbp]
\centering
\vspace{-15pt}
\subfigure[]{\label{fig:sixteena}\includegraphics[height=3.3cm]{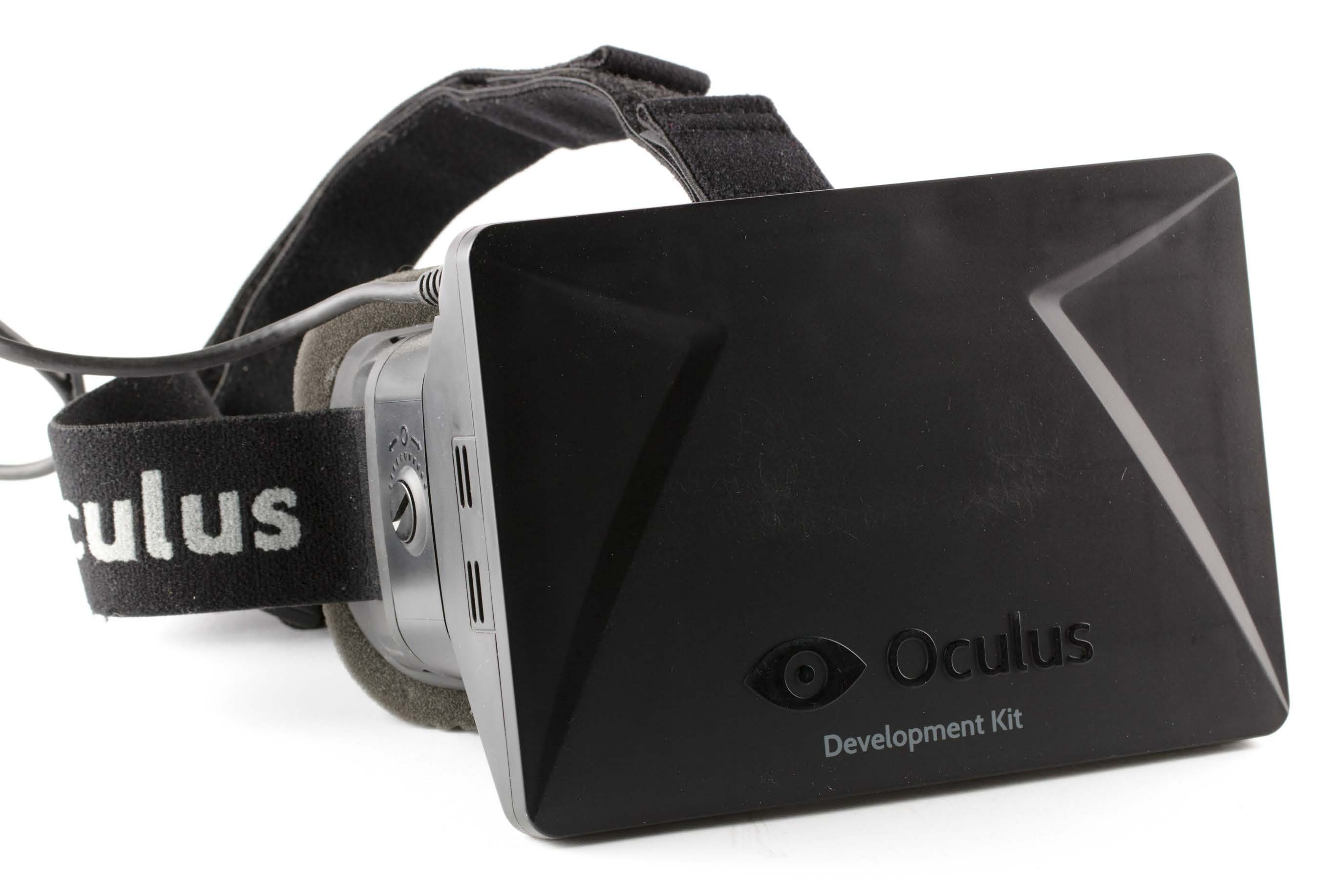}} \hfill 
\subfigure[]{\label{fig:sixteenb}\includegraphics[height=3.3cm]{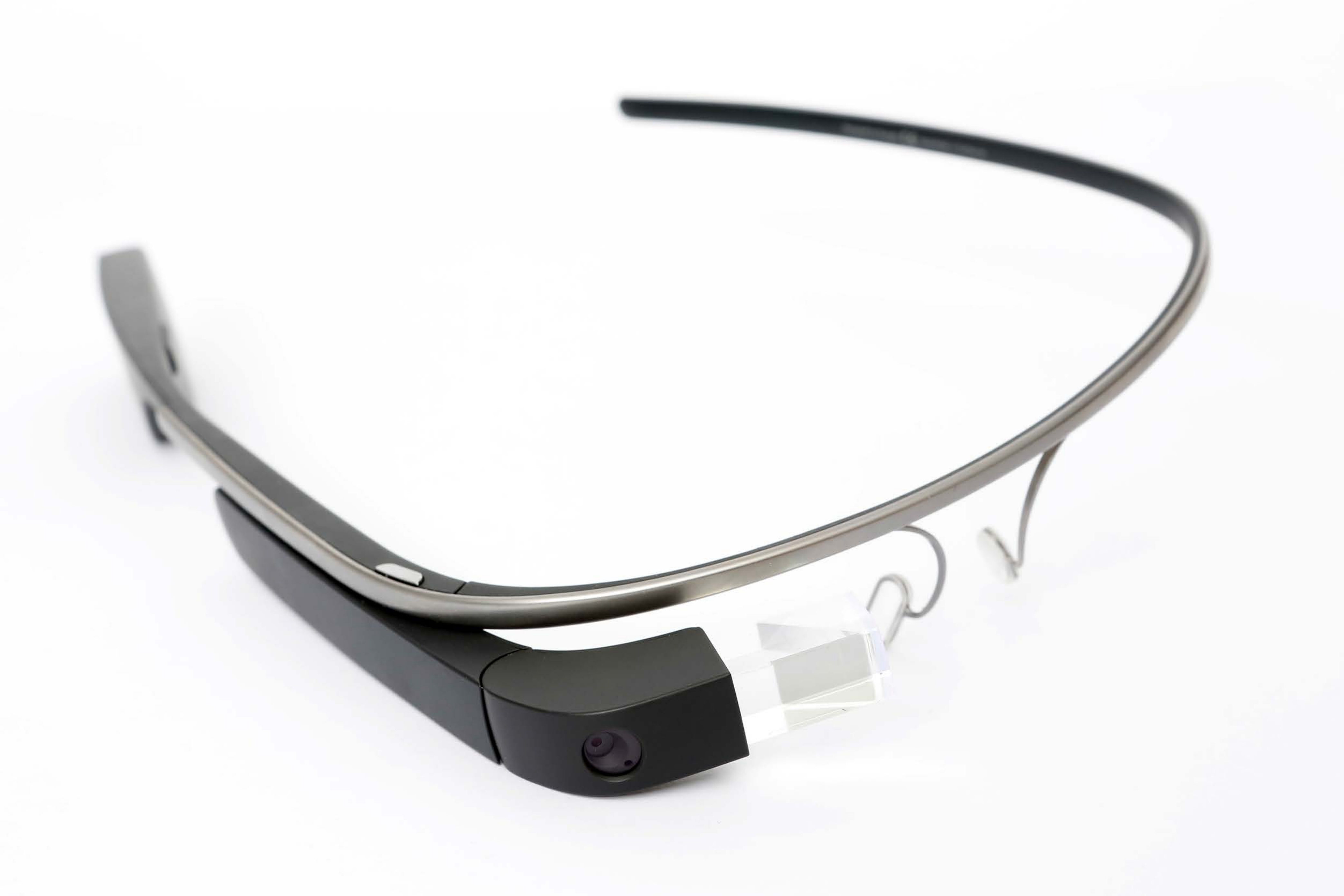}}
\subfigure[]{\label{fig:sixteenc}\includegraphics[height=3.3cm]{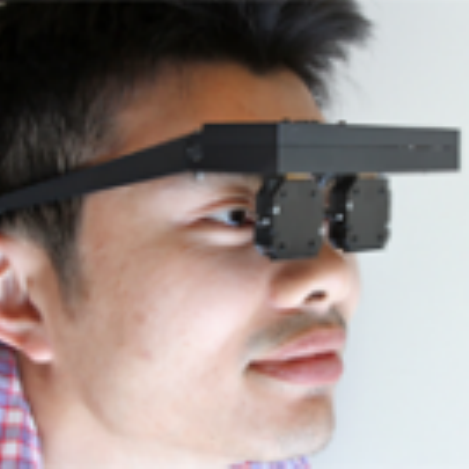}}
\vspace{-10pt}
\caption{ (a): \emph{Oculus Rift} developer edition. (b): Google Glass. (c): Near-eye light field project by NVidia.}
\label{fig:sixteen}
\end{figure}

\begin{wrapfigure}{L}{1.1cm}
	\vspace{-15pt}	
	\includegraphics[width=1.0cm]{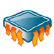}
	\vspace{-15pt}		
\end{wrapfigure} 
\noindent \textbf{Google Glass}, which was already announced as \emph{Project Glass} in 2012, becomes available through the explorer program in late 2013. and raises positive and negative attention, as well as concerns about privacy and ethical aspects (see Fig.\ref{fig:sixteenb}).

\vspace{0.1in}

\begin{wrapfigure}{L}{2.1cm}
	\vspace{-15pt}	
	\includegraphics[width=1.0cm]{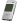}
	\includegraphics[width=1.0cm]{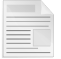}
	\vspace{-25pt}		
\end{wrapfigure} 
\noindent At ICRA, Li \etal present an amazing approach for \textbf{motion tracking with inertial sensors and a rolling-shutter camera} running in real-time on a mobile phone \cite{Li13}.

\vspace{0.1in}

\begin{wrapfigure}{L}{2.1cm}
	\vspace{-10pt}	
	\includegraphics[width=1.0cm]{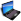}
	\includegraphics[width=1.0cm]{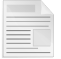}
	\vspace{-20pt}		
\end{wrapfigure} 
\noindent Tan \etal propose an approach to SLAM working in \textbf{dynamic environments}, allowing parts in the scene to be dynamic without breaking the mapping and tracking \cite{Tan13}.

\vspace{0.1in}

\begin{wrapfigure}{L}{1.2cm}
	\vspace{-0pt}	
	\includegraphics[width=1.2cm]{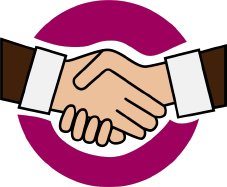}
	\vspace{-20pt}		
\end{wrapfigure} 
\noindent On November 24, 2013, \textbf{Apple Inc. confirms the purchase of PrimeSense} for about \$350 million \citelinks{AP13}. Primesense was working on shrinking their sensors to fit into mobiles at that point in time.

\vspace{0.1in}

\begin{wrapfigure}{L}{2.1cm}
	\vspace{-15pt}	
	\includegraphics[width=1.0cm]{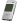}
	\includegraphics[width=1.0cm]{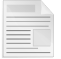}
	\vspace{-25pt}		
\end{wrapfigure} 
\noindent Taskanen \etal propose an approach to perform \textbf{full 3D reconstruction on a mobile monocular smartphone} and creating a dense 3D model with known absolute scale \cite{Taskanen13}.

\vspace{-5pt}
\section*{2014}

\begin{wrapfigure}{L}{1.2cm}
	\vspace{-10pt}	
	\includegraphics[width=1.2cm]{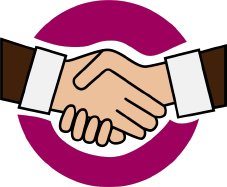}
	\vspace{-20pt}		
\end{wrapfigure} 
Three years after the acquisition, in January \textbf{Google sells Motorola Mobility} to Lenovo for \$2.91 million, however, keeping most of the patent portfolio \citelinks{GO14}.

\vspace{0.1in}

\begin{figure}[tbp]
\centering
\vspace{-5pt}
\subfigure[]{\label{fig:seventeena}\includegraphics[height=3.7cm]{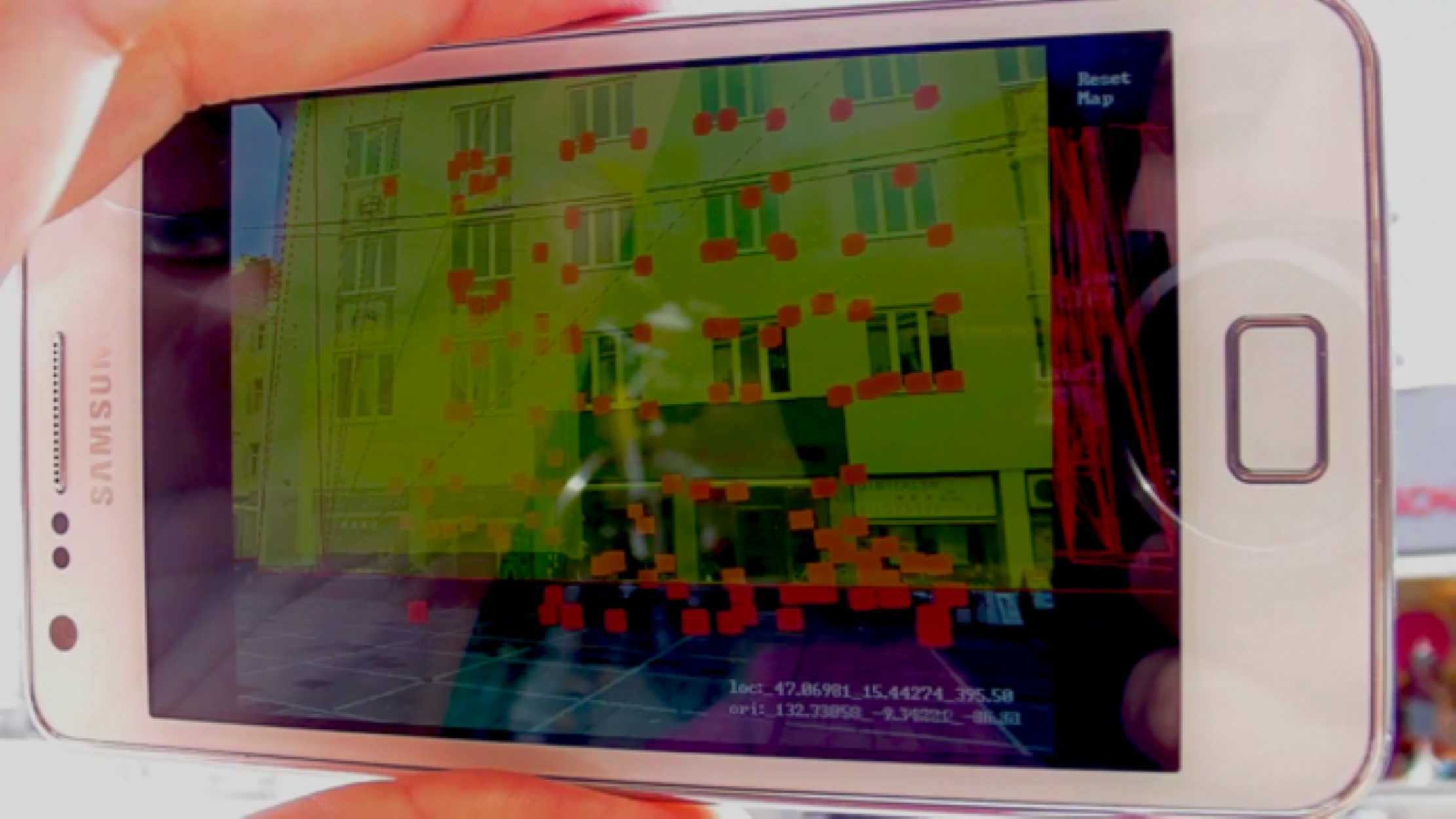}} \hfill 
\subfigure[]{\label{fig:seventeenb}\includegraphics[height=3.7cm]{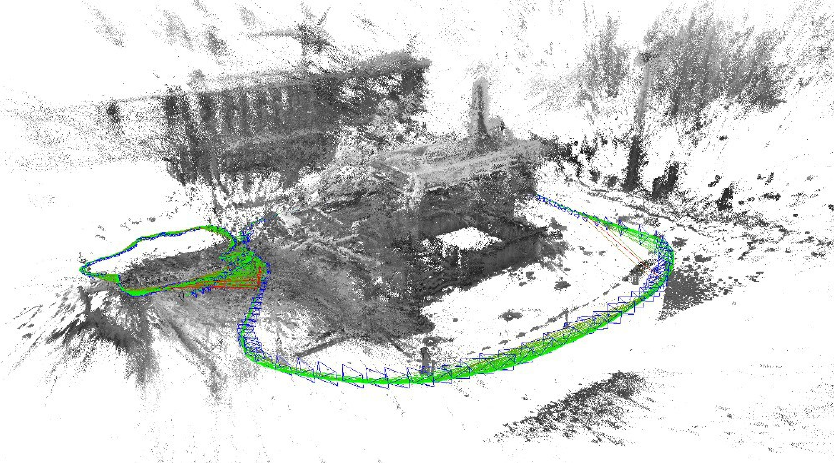}}
\vspace{-10pt}
\caption{ (a): SLAM map localization by Ventura \etal \cite{Ventura14}. (b): \emph{LSD-SLAM} reconstruction by Engel \etal \cite{engel14eccv}.}
\vspace{-10pt}
\label{fig:seventeen}
\end{figure}

\begin{wrapfigure}{L}{1.2cm}
	\vspace{-0pt}	
	\includegraphics[width=1.2cm]{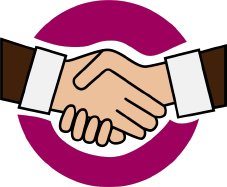}
	\vspace{-20pt}		
\end{wrapfigure} 
\noindent Also in January, \textbf{Qualcomm acquires Kooaba} \citelinks{QC14}, a Swiss ETH-spin-off founded in 2007, built around image recognition using SURF features. Kooaba's technology is integrated into the services provided through the Vuforia SDK.

\vspace{0.1in}

\begin{wrapfigure}{L}{1.1cm}
	\vspace{-10pt}	
	\includegraphics[width=1.0cm]{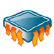}
	\vspace{-10pt}		
\end{wrapfigure} 
\noindent In February, Google announces \textbf{Project Tango} \citelinks{GO142}, which is an Android smartphone equipped with a full Kinect-like 3D sensor and hands out a few hundred units to developers and companies.

\vspace{0.1in}

\begin{wrapfigure}{L}{1.2cm}
	\vspace{-10pt}	
	\includegraphics[width=1.2cm]{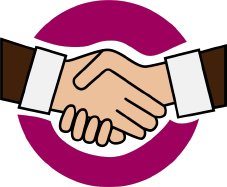}
	\vspace{-20pt}		
\end{wrapfigure} 
\noindent In March, \textbf{Facebook acquires Oculus VR} for \$2 billion, although Oculus does not make any consumer products at that point in time yet \citelinks{FB14}. This strengthens the hype in upcoming VR interfaces.

\vspace{0.1in}

\begin{wrapfigure}{L}{2.1cm}
	\vspace{-15pt}	
	\includegraphics[width=1.0cm]{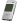}
	\includegraphics[width=1.0cm]{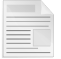}
	\vspace{-25pt}		
\end{wrapfigure} 
\noindent At VR, Ventura \etal present an approach to \textbf{localize SLAM maps built on a mobile phone} accurately wrt. a sparse 3D reconstruction of urban environments \cite{Ventura14} (see Fig.\ref{fig:seventeena}).

\vspace{0.1in}

\begin{wrapfigure}{L}{1.2cm}
	\vspace{-0pt}	
	\includegraphics[width=1.2cm]{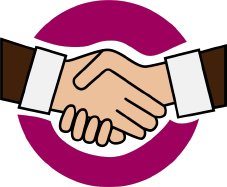}
	\vspace{-20pt}		
\end{wrapfigure} 
\noindent In April, \textbf{Microsoft} announces the \textbf{acquisition of Nokia's Devices and Services} unit for \$7.2 billion \citelinks{MS14}, as Nokia is the primary vendor for Windows devices devices, especially the Lumia phones.

\vspace{0.1in}

\begin{wrapfigure}{L}{3.2cm}
	\vspace{-15pt}	
	\includegraphics[width=1.0cm]{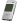}
	\includegraphics[width=1.0cm]{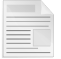}
	\includegraphics[width=1.0cm]{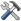}
	\vspace{-25pt}		
\end{wrapfigure} 
\noindent Following up on previous work at ICCV 2013 \cite{Engel13}, at ECCV Engel \etal present \textbf{LSD-SLAM}, a feature-less monocular SLAM algorithm using keyframes and semi-dense depth maps, and \textbf{release the code} to the public \cite{engel14eccv} (see Fig.\ref{fig:seventeenb}). At ISMAR, a mobile version is presented as well \cite{schoeps14ismar}.

\vspace{0.1in}

\begin{wrapfigure}{L}{2.1cm}
	\vspace{-10pt}	
	\includegraphics[width=1.0cm]{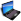}
	\includegraphics[width=1.0cm]{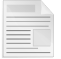}
	\vspace{-25pt}		
\end{wrapfigure} 
\noindent At 3DV, Herrera \etal present DT-SLAM \cite{Herrera14}. The key idea behind the approach is to \textbf{defer the triangulation step} of  2D features matched across keyframes until those have undergone a certain baseline, improving the overall robustness of SLAM.

\vspace{0.1in}

\begin{wrapfigure}{L}{2.1cm}
	\vspace{-10pt}	
	\includegraphics[width=1.0cm]{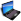}
	\includegraphics[width=1.0cm]{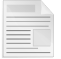}
	\vspace{-20pt}		
\end{wrapfigure} 
\noindent At ISMAR, Salas-Moreno \etal present \textbf{Dense Planar SLAM}, leveraging the assumption that many \textbf{man-made surfaces are planar} \cite{SalasMoreno14}.

\vspace{-5pt}
\section*{2015}

\begin{wrapfigure}{L}{1.2cm}
	\vspace{-15pt}	
	\includegraphics[width=1.2cm]{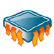}
	\vspace{-20pt}		
\end{wrapfigure} 
In January, Microsoft announces the \emph{Hololens}, a headset to fuse AR and VR \citelinks{MS15} to be made available later in 2015. The device is a complete computer with a see-through display and several sensors. 

\vspace{0.1in}

\begin{wrapfigure}{L}{1.2cm}
	\vspace{-10pt}	
	\includegraphics[width=1.2cm]{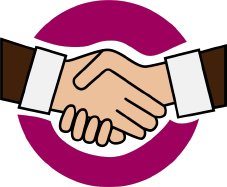}
	\vspace{-20pt}		
\end{wrapfigure} 
\noindent In May, \textbf{DAQRI}, a company working on AR helmets, \textbf{acquires ARToolworks} \citelinks{ART15}. \textbf{Oculus VR} announced the \textbf{acquisition of Surreal Vision}, bringing the company’s expertise on recreating real-time 3D representations of the outside world into virtualized environments \citelinks{OC15}. A few days later, the \textbf{acquisition of} German AR company \textbf{Metaio by Apple} is announced \citelinks{AP15}. Metaio was a major player in distribution of AR technology to developers through their SDK, which ends abruptly now. 

\vspace{0.1in}

\begin{wrapfigure}{L}{2.2cm}
	\vspace{-15pt}	
	\includegraphics[width=1.0cm]{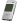}
	\includegraphics[width=1.0cm]{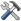}
	\vspace{-20pt}		
\end{wrapfigure} 
\noindent In June, \textbf{Magic Leap} announces that they will \textbf{release its AR platform SDK} to the public soon. It is expected to support Unity and the Unreal engine \citelinks{ML15}.

\vspace{0.1in}

\begin{wrapfigure}{L}{1.1cm}
	\vspace{-10pt}	
	\includegraphics[width=1.0cm]{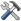}
	\vspace{-15pt}		
\end{wrapfigure}
\noindent HTC and Valve start shipping their developer hardware in very limited quantities \citelinks{VA15}, including a Vive headset and two Lighthouse base stations as passive components together with the two wireless  Steam VR controllers.

\vspace{0.3in}

\begin{wrapfigure}{L}{1.2cm}
	\vspace{5pt}	
	\includegraphics[width=1.2cm]{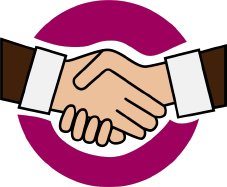}
	\vspace{-20pt}		
\end{wrapfigure} 
\noindent In October 2015, Qualcomm sells its business unit in Vienna responsible for the development of the Augmented Reality SDK \emph{Vuforia} to PTC \citelinks{PT15} for 65 Mio. USD. 

\vspace{0.20in}

\begin{wrapfigure}{L}{2.2cm}	
	\vspace{-15pt}	
	\includegraphics[width=1.0cm]{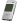}
	\includegraphics[width=1.0cm]{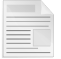}
	\vspace{-10pt}		
\end{wrapfigure} 
\noindent At ISMAR 2015, the group of Graz University of Technology wins the best paper award for their work "Instant Outdoor Localization and {SLAM} Initialization from 2.5D Maps" \cite{ArthPVSL15}. The algorithm uses OpenStreetMap data and calculates the pose for an image based on building edges and semantic image information. 

\section*{Acknowledgements}

Thanks go to the ISMAR09 mobile committee and all others for their valuable suggestions.

% Bibliography.
% -------------

\parskip=0pt
\parsep=0pt

\bibliographystyle{ieeetrsrt}
\bibliography{historyAR}

\parskip=0pt
\parsep=0pt

\bibliographystylelinks{plain}
\bibliographylinks{links}

\end{document}